\documentclass{article}

\usepackage[labelfont=bf]{caption} 
\usepackage[font={small}]{caption}  

\usepackage{graphicx}
\usepackage{dcolumn}
\usepackage{bm}
\usepackage[pdftex,colorlinks=true,linkcolor=blue,citecolor=blue]{hyperref}
\usepackage[utf8]{inputenc}
\usepackage{lipsum}  

\usepackage{amsmath}
\usepackage{authblk}
\usepackage{amsfonts}
\usepackage{placeins}

\usepackage[numbers, sort&compress]{natbib} 
\usepackage{bibunits}
\usepackage{cancel}
\usepackage{steinmetz}
\usepackage{ bbold }

    \usepackage{lineno}   
    \usepackage{amsmath}  
    \usepackage{etoolbox}

    \newcommand*\linenomathpatch[1]{%
    \cspreto{#1}{\linenomath}%
    \cspreto{#1*}{\linenomath}%
    \csappto{end#1}{\endlinenomath}%
    \csappto{end#1*}{\endlinenomath}%
    }

    \linenomathpatch{equation}
    \linenomathpatch{gather}
    \linenomathpatch{multline}
    \linenomathpatch{align}
    \linenomathpatch{alignat}
    \linenomathpatch{flalign}

\usepackage{amssymb}
\usepackage{pdflscape}

    \usepackage[table, usenames, dvipsnames]{xcolor}
    \usepackage{makecell}
    \usepackage{boldline}
    \setcellgapes{3pt}

\usepackage{textgreek}

\usepackage[strict]{changepage} 
\usepackage{nicefrac} 

\defaultbibliographystyle{naturemag}
\defaultbibliography{thesisrefsol}

    \usepackage{soul}   
    \setstcolor{red} 
    \setul{}{1.5pt}

\title{Electrically-Controlled Suppression of Rayleigh Backscattering in an Integrated Photonic Circuit}

\author{
    Oğulcan E. Örsel $^1$, Jiho Noh $^2$ , Gaurav Bahl $^{2\ast}$ \\
    $^1$ Department of Electrical $\&$ Computer Engineering, \\
    $^2$ Department of Mechanical Science $\&$ Engineering, \\
    University of Illinois at Urbana–Champaign, Urbana, IL 61801, USA \\
    \vspace{4pt}
    $^\ast$ bahl@illinois.edu
}

\date{}

\begin{document}
\begin{bibunit}

\maketitle

\begin{abstract}

Undesirable light scattering is an important fundamental cause for photon loss in nanophotonics. Rayleigh backscattering can be particularly difficult to avoid in wave-guiding systems and arises from both material defects and geometric defects at the subwavelength scale. It has been previously shown that systems with broken time-reversal symmetry (TRS) can naturally suppress detrimental Rayleigh backscattering, but these approaches have never been demonstrated in integrated photonics or through practical TRS-breaking techniques. In this work, we show that it is possible to suppress disorder-induced Rayleigh backscattering in integrated photonics via electrical excitation, even when defects are clearly present. Our experiment is performed in a lithium niobate on insulator (LNOI) integrated ring resonator at telecom wavelength, in which TRS is strongly broken through an acousto-optic interaction that is induced via radiofrequency input. We present evidence that Rayleigh backscattering in the resonator is almost completely suppressed by measuring both the optical density of states and through direct measurements of the back-scattered light. We additionally provide an intuitive argument to show that, in an appropriate frame of reference, the suppression of backscattering can be readily understood as a form of topological protection.

\end{abstract}

\maketitle

\begin{figure}[htp]
    \begin{adjustwidth*}{-1in}{-1in}
    \hsize=\linewidth
    \includegraphics[width=1.1\textwidth]{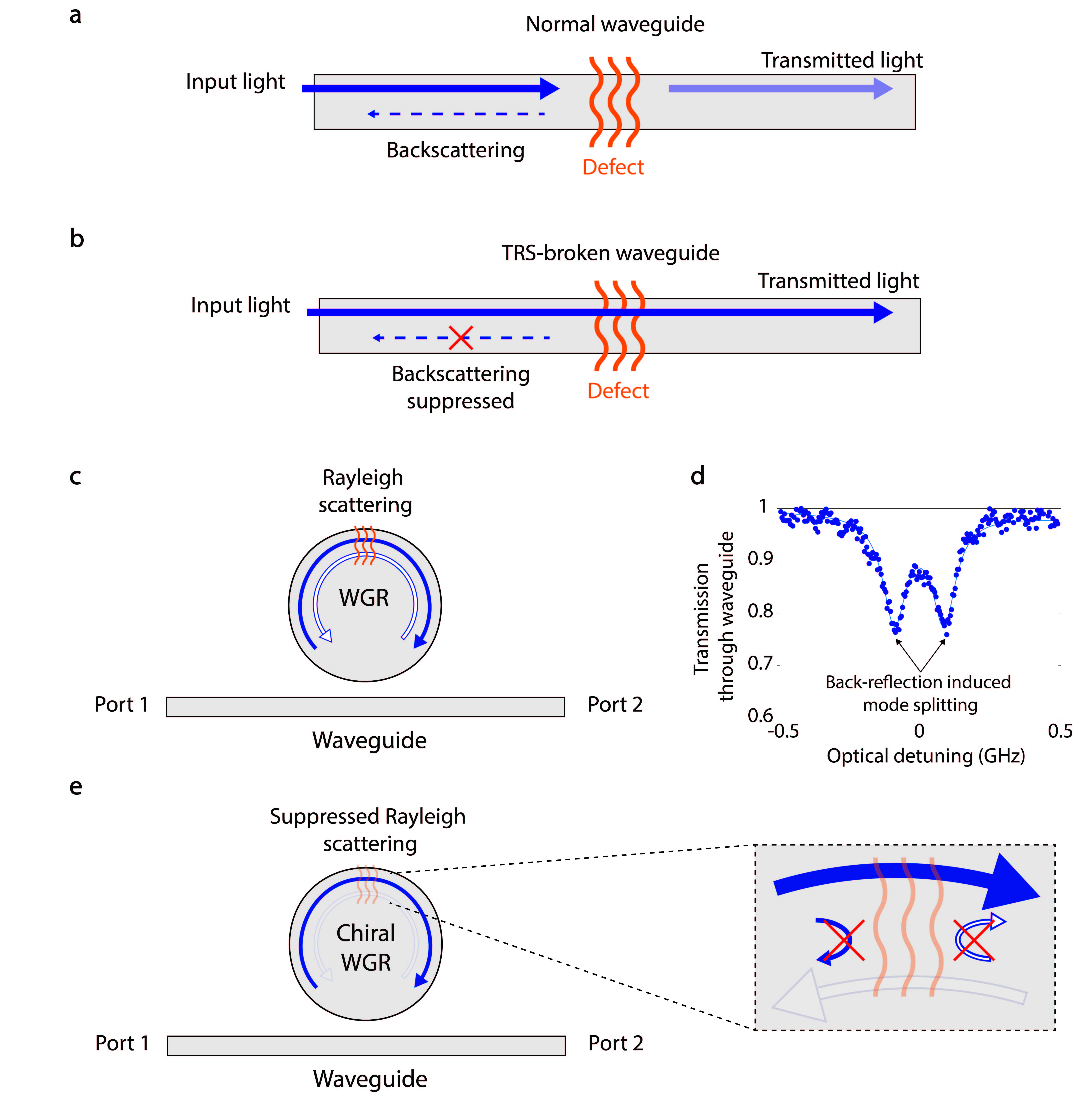}
    \centering
    \caption{
        \textbf{Backscattering suppression via chiral dispersion engineering.}
        \textbf{(a)} A 1D waveguide supports symmetric forward and backward propagating optical modes. Subwavelength defects can produce undesirable Rayleigh backscattering that reduces forward transmission.
        \textbf{(b)} In a waveguide with suitably broken time-reversal symmetry -- i.e. a chiral waveguide -- there may not be any back-propagating states available, which suppresses the backscattering from taking place in spite of the presence of the defect.
        \textbf{(c)} Similarly, whispering gallery resonators (WGRs) are quasi-1D systems that support nominally degenerate cw and ccw optical modes. Rayleigh backscattering within the resonator can couple these counter-propagating modes leading to the loss of their distinguishable directionality.
        \textbf{(d)} Experimental example of a doublet of cw/ccw hybridized modes induced by Rayleigh scattering in a integrated LNOI WGR shown in later figures.
        \textbf{(e)} In a chiral WGR, the optical density of states for cw and ccw circulation can be different, which as we show can suppress the Rayleigh backscattering.
    }
    \label{fig:1}
    \end{adjustwidth*}
\end{figure}

Defects are unavoidable in almost every optical device~\cite{Weiss_Sandoghdar_Hare_Lefevre-Seguin_Raimond_Haroche_1995, Gorodetsky_Pryamikov_Ilchenko_2000}, and especially in integrated photonics~\cite{Borselli_Johnson_Painter_2005, Morichetti_Canciamilla_Ferrari_Torregiani_Melloni_Martinelli_2010,Liu_Bruch_Gong_Lu_Surya_Zhang_Wang_Yan_Tang_2018} due to the limitations of crystal growth and nanofabrication methods. 
Common subwavelength defects like intrinsic material stresses, density variation, and surface roughness can lead to Rayleigh scattering which can result in extra propagation loss, limitations on optical Q-factors, and undesirable inter-modal conversion \cite{Marcuse_1969,Mazzei_Gotzinger_de-S_Menezes_Zumofen_Benson_Sandoghdar_2007,Kippenberg_Tchebotareva_Kalkman_Polman_Vahala_2009}. 
In particular, disorder-induced Rayleigh backscattering ({Fig.~\ref{fig:1}a}) is a common problem in waveguiding structures that has been directly linked to stability problems in frequency combs \cite{Griffith_Lau_Cardenas_Okawachi_Mohanty_Fain_Lee_Yu_Phare_Poitras_etal, Suh_Yang_Yang_Yi_Vahala_2016}, elimination of directional gain in micro-ring lasers \cite{Kim_Kwon_Shim_Jung_Yu_2014}, and increased bit error rates in integrated modulators~\cite{Bahadori_Rumley_Cheng_Bergman_2018}. While isolators and circulators can be used to navigate such backscattering, they only serve a fixative function but do not mitigate the scattering problem itself. The photon directionality is not preserved, forward propagating power reduces, and coupling into undesirable modes still occurs.

One approach to prevent disorder-induced backscattering is to strive for a perfect disorderless optical system, that is, by reducing or eliminating scatterersfrom the medium. However, since no fabrication method can produce truly defect-free optics, and sometimes defects can be introduced during operation or appear dynamically due to thermal fluctuations, this approach is not entirely feasible. A better alternative is to simply break time-reversal symmetry (TRS) in the medium so that modes propagating in opposite directions are not symmetric in frequency-momentum space. The presence of such a substantial contrast in the density of states for counter-propagating modes -- often termed as chiral dispersion -- can suppress elastic backscattering events from occurring at all (Fig.~\ref{fig:1}b).
Chiral dispersion naturally occurs in magneto-optic materials and the inhibition of coherent backscattering has indeed been experimentally observed \cite{Lenke_Maret_2000}, but this is not a convenient option for integrated photonics due to foundry-incompatible materials and the need for magnetic biasing. 
Chiral dispersions are also possible in 2D Fermionic systems, e.g., via the quantum Hall effect (QHE) \cite{Buttiker_1988}. Optical analogues of the QHE can also be realized in photonic topological insulator (PTI) metamaterials, using both gyromagnetic effects in photonic crystals \cite{Haldane_Raghu_2008, Wang_Chong_Joannopoulos_Soljacic_2009} and Floquet engineering \cite{Rechtsman_Zeuner_Plotnik_Lumer_Podolsky_Dreisow_Nolte_Segev_Szameit_2013}, 
but these approaches are not currently practical for integrated photonics.
Importantly, it was also recently confirmed~\cite{Rosiek_Stobbe_2023} that topological protection only applies to defects that do not violate the underlying protective symmetries of the photonic topological insulator, and that scattering from the more practically encountered subwavelength defects may not be avoidable unless TRS is explicitly broken such as in the PTI examples cited above.
In previous studies we have demonstrated an alternative approach to suppress disorder-induced backscattering in common optical dielectrics, without needing any special micro-structuring, via TRS-breaking with photon-phonon interactions \cite{Kim2017, Kim_Taylor_Bahl_2019}. However, this method is also generally inconvenient for photonic integrated circuits as it requires optical pumping.

In this work we demonstrate the near-complete suppression of disorder-induced backscattering through electrically-controlled acoustic pumping in an integrated photonic system at telecom wavelength.
For the demonstration of this effect we consider high Q-factor whispering gallery type resonators (WGRs), such as those with ring or racetrack geometry, in which backscattering can often be observed from subwavelength defects. A WGR is essentially a 1D wave-guiding system with periodic boundary conditions that supports counter-propagating optical modes that are frequency degenerate, nominally orthogonal, and form momentum-reversed pairs. Subwavelength scatterers can induce coupling between such counter-propagating mode pairs (see Fig.~\ref{fig:1}c) causing the optical states to be disturbed from their intrinsic state in both directions. For a low backscattering rate the modes simply broaden from their intrinsic linewidth \cite{Mazzei_Gotzinger_de-S_Menezes_Zumofen_Benson_Sandoghdar_2007}. With higher backscattering rates (i.e., more than the loss rate of the optical mode) the modes exhibit distinct doublet characteristics as shown in Fig.~\ref{fig:1}d, which is a clear signature of Rayleigh backscattering \cite{Weiss_Sandoghdar_Hare_Lefevre-Seguin_Raimond_Haroche_1995, Gorodetsky_Pryamikov_Ilchenko_2000,Borselli_Johnson_Painter_2005, Mazzei_Gotzinger_de-S_Menezes_Zumofen_Benson_Sandoghdar_2007, Puckett_Liu_Chauhan_Zhao_Jin_Cheng_Wu_Behunin_Rakich_Nelson_2021, Ji:17}. 
If we are able to dynamically induce a strong chiral dispersion within this resonator by breaking TRS ({Fig.~\ref{fig:1}e}), the backscattering should be suppressed, removing the doublet and restoring the original modes with an improved quality factor.
As we will show, we can experimentally achieve chiral dispersion in integrated lithium niobate WGRs through a very practical acousto-optic TRS-breaking technique~\cite{Sohn18,Sohn_Orsel_Bahl_2021}, and confirm the suppression of disorder-induced backscattering internal to the device.

\vspace{12pt}   
    
\begin{figure}[htp]
    \begin{adjustwidth*}{-1in}{-1in}
    \hsize=\linewidth
    \includegraphics[width=1.3\textwidth]{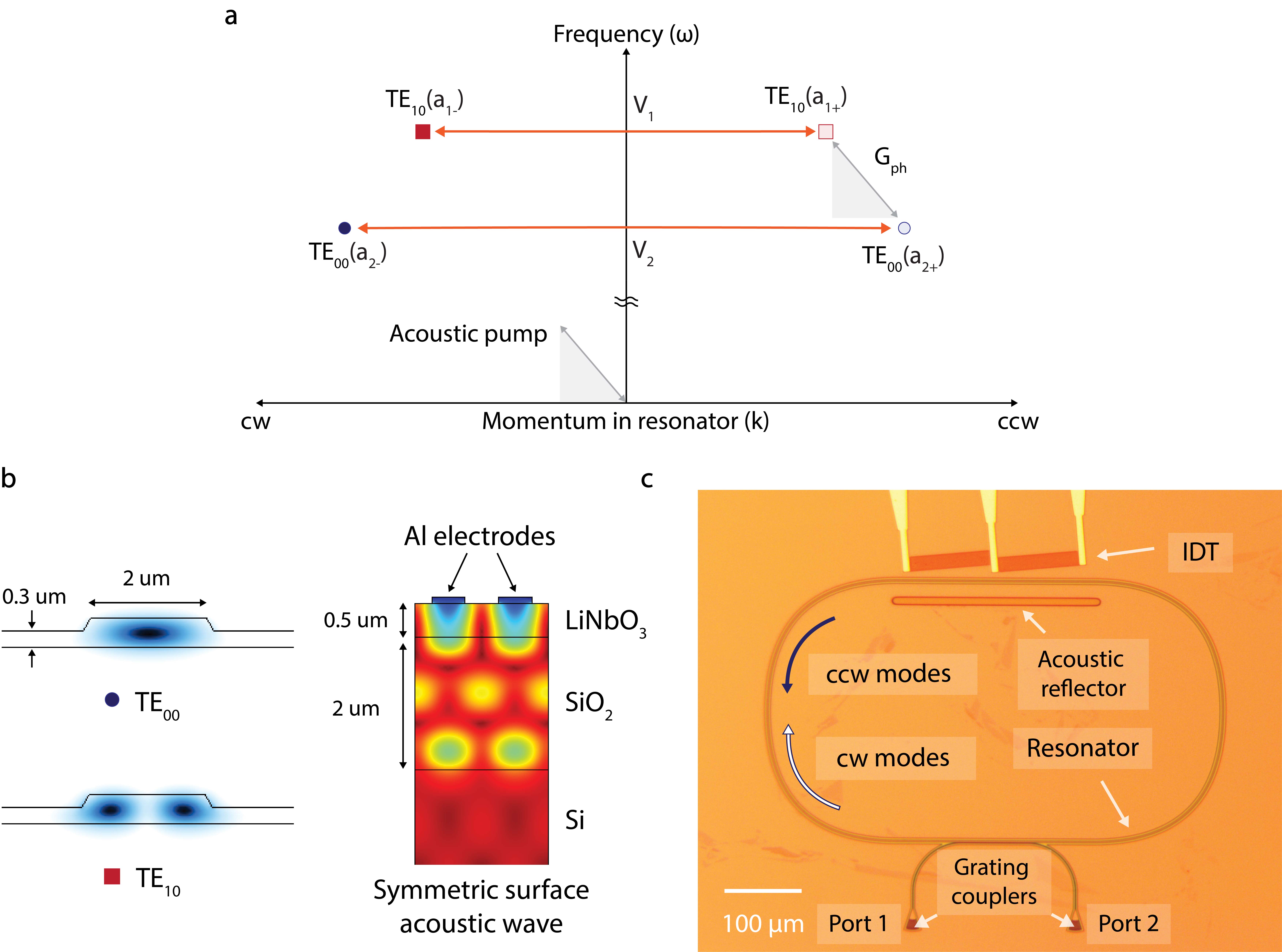}
    \centering
    \caption{
        \textbf{Implementation of chiral dispersion in LNOI racetrack resonators.} 
        \textbf{(a)} This frequency-momentum ($\omega-k$) diagram shows the modes of the racetrack resonator in the presence of direction-sensitive acousto-optic interaction ($G_{ph}$) and also disorder-induced backscattering ($V_1, V_2$). The relative placement of optical modes in $\omega-k$ space matches the experimental devices presented in this work. It can be seen that, when the acoustic pump has momentum in the cw direction, the phase matching for acousto-optic scattering is only satisfied for counterclockwise (ccw) circulation around the racetrack. 
        \textbf{(b)} The resonator is designed to support TE\textsubscript{00} and TE\textsubscript{10} optical modes as simulated here. 
        We also simulate the symmetric surface acoustic wave at the location of the interdigital transducer (IDT) which is indicated by the aluminum (Al) electrodes.
        \textbf{(c)} A photograph of the fabricated LiNbO\textsubscript{3} device is presented. The IDT is angled to produce acoustic wave momentum along the clockwise (cw) circulation direction while an acoustic reflector helps to form a transverse standing wave.
        Some e-beam resist residue associated with fabrication is apparent on the surface. 
        Port 1 and Port 2 indicate the grating couplers by which light is coupled into the waveguide.
    }
    \label{fig:2}
    \end{adjustwidth*}
\end{figure}

 Our underlying approach to breaking TRS in WGRs has been previously described in~\cite{Sohn18,Sohn_Orsel_Bahl_2021}. As a brief recap, the configuration of optical modes within our example WGR is presented in {Fig.~\ref{fig:2}a}. The resonator supports both TE\textsubscript{00} and TE\textsubscript{10} modes, although other modes may be used, with the undesirable scattering between cw and ccw circulations introduced as coupling terms $V_1$ and $V_2$. We ensure that these optical modes are close together in frequency space but have large separation in momentum space. Propagating phonons having frequency and momentum that bridge the gap between these optical modes can be introduced via piezoelectric excitation. The direction of the phonon propagation sets which circulation of optical modes are hybridized through acousto-optic scattering and breaks symmetry in the system. Aside, we note that there is also a necessary index texture required to ensure a non-zero overlap integral for the scattering process, the details of which can be found in~\cite{Sohn18,Sohn_Orsel_Bahl_2021} and in the Supplement \S\ref{sec:Modelling}. With sufficiently large acousto-optic scattering rate $G_{ph}$ the optical modes hybridize only for one circulation direction around the device, leading to a very strong directional contrast in the optical density of states and reduced spectral overlap.

We developed our telecom wavelength ($\sim$1550 nm) experimental demonstration on a LiNbO\textsubscript{3}-on-insulator (LNOI) integrated photonics platform since the large piezoelectric coefficients of LiNbO\textsubscript{3} enable the efficient excitation of surface acoustic waves and produce a significant acousto-optic coupling rate.
The LNOI platform is also convenient for realizing integrated WGRs with relatively large optical Q-factors ($>10^6$). This combination of high optical Q-factors and large acousto-optic coupling rates promotes the realization of strong mode hybridization in our optical devices, thus enabling chiral dispersion as demonstrated in \cite{Sohn_Orsel_Bahl_2021}.
As described above, we design and fabricate (Fig.~\ref{fig:2}c) a racetrack resonator supporting TE\textsubscript{10} and TE\textsubscript{00} modes (Fig.~\ref{fig:2} b,c) which can act as a two-level photonic molecule. We access the resonator with a single-mode waveguide in shunt configuration and add grating couplers to each end of this waveguide to provide off-chip optical access to the device. 
An aluminum interdigital transducer (IDT) is co-fabricated with orientation along the Y-$30^{\circ}$ direction of the lithium niobate crystal~\cite{Kuznetsova_Zaitsev_Joshi_Borodina_2001} to maximize the electromechanical coupling rate to the symmetric surface acoustic mode (Fig.~\ref{fig:2}b). The phonons produced by the actuator propagate along a portion of the racetrack with momentum corresponding to the cw circulation direction. The actuator design is detailed in \cite{Sohn_Orsel_Bahl_2021} and includes an acoustic reflector to produce a transverse standing wave for producing the necessary acoustic wave texture.
The resulting acousto-optic coupling in this photonic molecule produces an effect analogous to an Autler-Townes splitting phenomenon \cite{Sohn_Orsel_Bahl_2021}.
The optical ports are explicitly labeled $1,2$ in {Fig.~\ref{fig:2}c} such that the $S_{21}$ ($S_{12}$) measurement through the waveguide reads the ccw (cw) optical states of the resonator.
Importantly, the acousto-optic coupling is phase matched only in one direction, i.e., for the selected modes it is engaged only in optical S{\textsubscript{21}} transmission measurements from port 1 to port 2 corresponding to the ccw circulation direction around the resonator.

\begin{figure}[htp]
    \begin{adjustwidth*}{-1in}{-1in} 
    \hsize=\linewidth
    \includegraphics[width=1.25\textwidth]{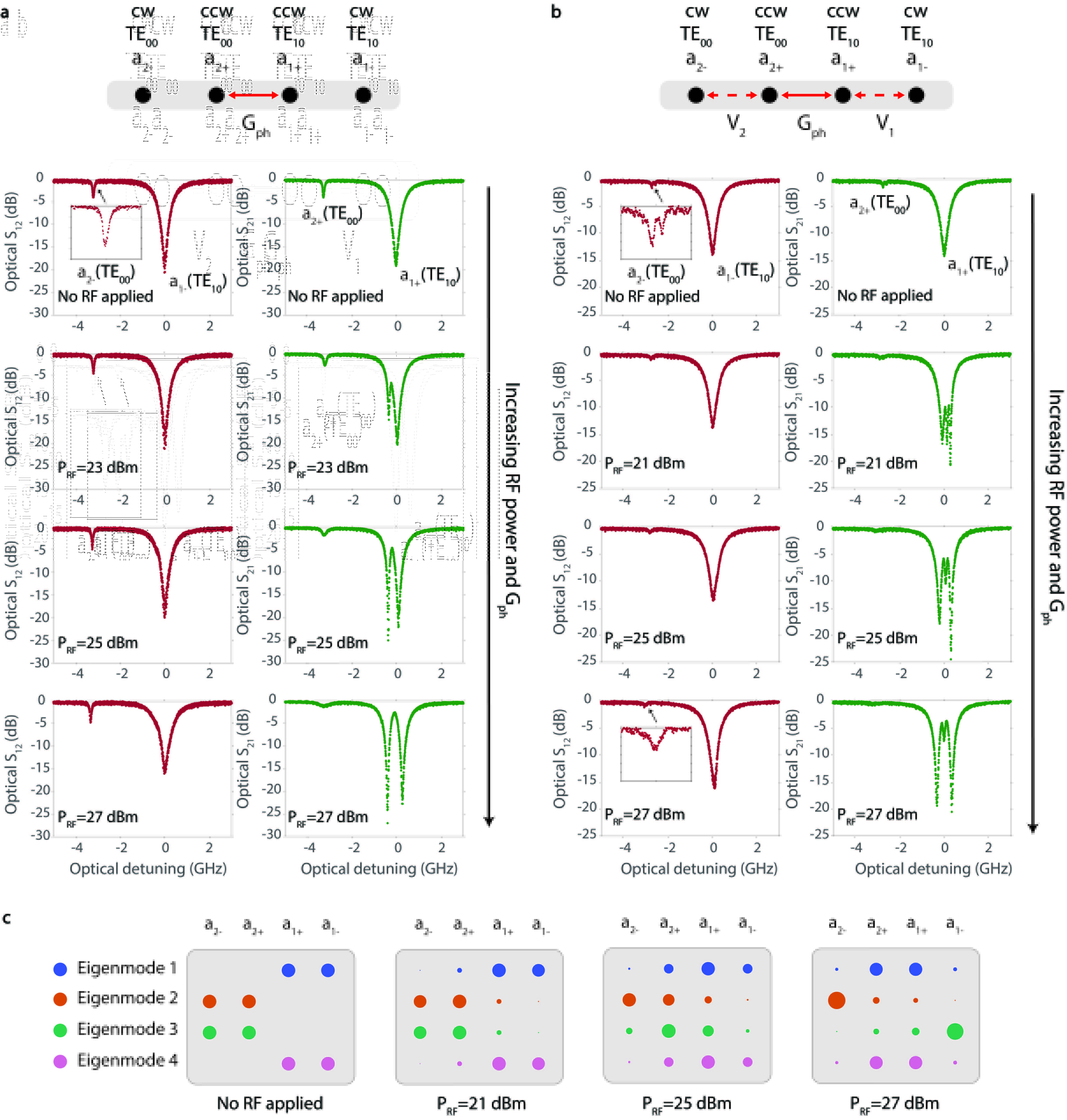}
    \centering
    \caption{
    \textbf{Experimental demonstration of phonon-induced chiral dispersion and the localization of cw modes.}
    \textbf{(a,b)} Measured through-waveguide transmission (S-parameters $S_{21}, S_{12}$ are described in the text) for devices having (a) negligible backscattering and (b) significant backscattering. The insets show the zoomed-in versions of the TE\textsubscript{00} modes in the cw circulation direction. Numerical quantification of the linewidths and scattering rates is presented in Supplementary Table~\ref{tab:DeviceParameters}. In the top row we present a 1D chain equivalent of these two cases with mapping as described in the Supplement \S{\ref{sec:DerivationHamiltonian}}.
    The key observation of the suppression of backscattering is made in the insets of (b), where we observe the cw TE{\textsubscript{00}} mode recover from a doublet into a singlet state when a TRS-breaking cw RF stimulus is provided to the IDT.
    \textbf{(c)} Simulated eigenvectors for the device in Fig.~\ref{fig:3}b under a rotating frame of reference described in \S{\ref{sec:DerivationHamiltonian}}.
    With increasing acousto-optic coupling $G_{ph}$ we observe a topological transition that leads to the emergence of protected boundary modes (Eigenmode 2,3) as explained in the manuscript.
    }
    \label{fig:3}
    \end{adjustwidth*}
\end{figure}


In Fig.~\ref{fig:3}, we present our experimental results for two devices that show both negligible (Fig.~\ref{fig:3}a) and significant (Fig.~\ref{fig:3}b) backscattering. The backscattering is unintentional and its source is not definitive, although it is speculated to originate either from device geometry defects or from the e-beam resist residue visible on the surface.
We use a heterodyne detection system (Supplementary Fig.~\ref{supfig:2}) to probe the optical signal transmissions. With no RF power applied (top data row of Fig.~\ref{fig:3}a,b), we observe that both optical transmission spectra show two distinct dips corresponding to the TE\textsubscript{00} and TE\textsubscript{10} modes of the resonator. The TE\textsubscript{10} mode shows a more significant dip than the TE\textsubscript{00} mode due to a larger evanescent field that increases coupling to the waveguide.
The distinction in backscattering rate can be experimentally observed from the top insets in Fig~\ref{fig:3}a,b. The device with significant backscattering (Fig~\ref{fig:3}b) shows the characteristic doublet for the TE\textsubscript{00} mode, which is in stark contrast to the case with negligible backscattering (Fig~\ref{fig:3}a).
Due to the larger intrinsic loss rate, backscattering occurring within the TE\textsubscript{10} mode is not manifested as a direct mode splitting but is instead revealed as some amount of linewidth broadening and an additional back-reflection from the resonator (see Supplement \S\ref{sec:Reflectioncalibration}).
We can now introduce photon-photon coupling in both systems by electrically launching phonons via the co-fabricated IDT. We specifically use a surface acoustic wave near 3 GHz, as simulated in Fig~\ref{fig:2}b, since its frequency and momentum best match the $\omega-k$ spacing of the TE\textsubscript{00} and TE\textsubscript{10} modes, and its shape provides the best overlap integral for acousto-optic scattering. A discussion of the transducer RF reflection spectrum is presented in the Supplement \S\ref{sec:RF-S-Parameters}.

In Fig.~\ref{fig:3}a, we show the evolution of the optical transmission through the device
having negligible backscattering as a function of increasing RF input power (i.e., increasing acousto-optic coupling rate). In the phase-matched direction (the optical S\textsubscript{21}), the optical modes significantly hybridize, and as a result the TE\textsubscript{10} mode shows a frequency splitting behavior proportional to the acousto-optic coupling rate. On the other hand, the optical modes remain unperturbed in the non-phase matched direction (the optical S\textsubscript{12}), thus demonstrating the chiral dispersion we wish to utilize.

We now look more closely at the device with higher backscattering, shown in Fig~\ref{fig:3}b. The evolution of the TE\textsubscript{10} mode pair (i.e., cw and ccw) 
with increasing acousto-optic coupling shows that, even as the mode experiences a $G_{ph}$-induced splitting in the phase (ccw) matched direction, a smaller third state can be seen peeking through at the original frequency. This extra state is apparent due to the disorder-induced backscattering ($V_2$) into the ccw TE\textsubscript{10} mode. A detailed analysis is presented in Supplement {\S\ref{sec:Modelling}, Eqn.~\ref{eq:three_mode}}. 
This extra state diminishes as the acousto-optic coupling $G_{ph}$ surpasses the backscattering rates $V_1, V_2$, and should eventually disappear since the cw and ccw modes are no longer spectrally overlapping.

At this point it is informative to consider the topological configuration of the system of modes, by unwrapping $\omega-k$ diagram of Fig.~\ref{fig:2}a into a flattened 1D chain. We can accomplish this using a frame of reference rotating with the acoustic pump frequency $\Omega$ 
(see Supplement \S\ref{sec:DerivationHamiltonian}), producing the 1D chains shown in Fig.~\ref{fig:3}a,b. 
We can then compute the eigenvectors of the system, as shown in Fig.~\ref{fig:3}c, to evaluate how the modes will evolve at various RF/acoustic pump powers.
With no acousto-optic coupling the spatial profiles of all four eigenvectors show only the hybridization between forward and backward modes.
With increased RF power, however, the forward and backward modes begin to diffuse to all other sites (see cases for 21 dBm and 25 dBm).
At very large RF power (the 27 dBm case) we observe significant localization of the backward propagating modes. 
This phenomenon can be readily understood in terms of the topological Su-Schrieffer-Heeger (SSH) model for dimerized chains~\cite{Su_Schrieffer_Heeger_1979}.
Here the backscattering $V_1, V_2$ takes the role of the intra-dimer bond while the acousto-optical modulation $G_{ph}$ takes the role of the inter-dimer bond.
In this analogy, familiar readers will see how exponentially localized edge states must emerge (labeled as Eigenmodes 2 and 3) when $G_{ph} \gg V_1, V_2$ since the chain transforms into the topologically non-trivial phase.
A more detailed discussion on the analogy to the SSH model and possible extension to a longer SSH chain can be found in the Supplement \S\ref{sec:SimilaritiesSSH}.
Experimentally, we can confirm this effect when observing the ccw TE\textsubscript{00} mode labeled $a_{2-}$ in the S\textsubscript{12} spectrum -- the inset shows clearly that the mode has recovered to a simple Lorentzian indicating also that the backscattering has been inhibited.

\begin{figure}[htp]
    \begin{adjustwidth*}{-1in}{-1in} 
    \hsize =\linewidth
        \includegraphics[width=1.3\textwidth, clip=true, trim=0in 0.5in 0in 0in]{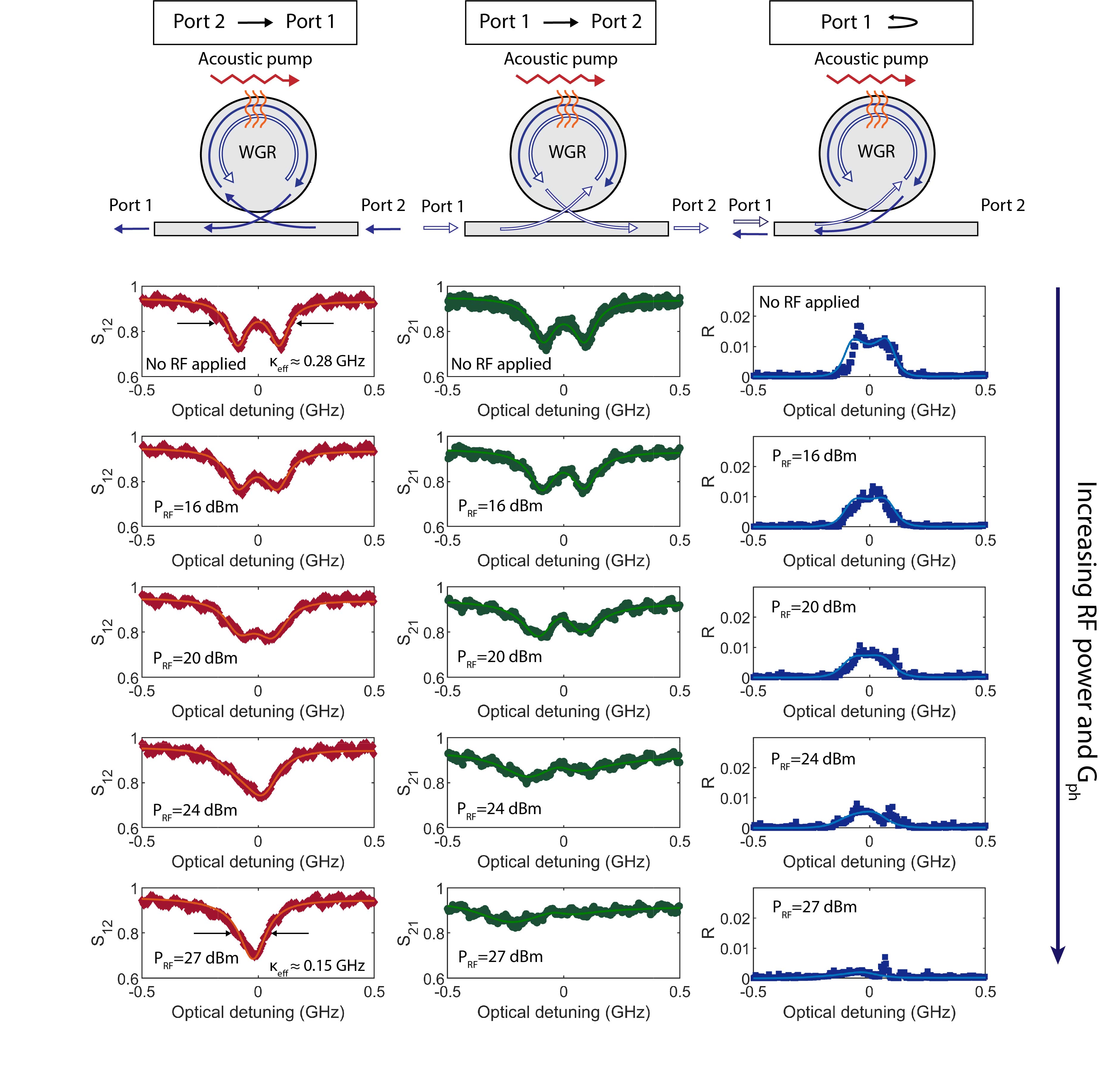}
    \centering
    \caption{
    \textbf{Experimental demonstration of the recovery of the TE\textsubscript{00} mode and elimination of the disorder-induced reflection.}
    The experimentally measured optical S parameters are presented (reflection coefficient $R = $ S\textsubscript{11}$=$ S\textsubscript{22}) near the TE\textsubscript{00} mode for a device having high rate of disorder-induced backscattering (all relevant device parameters can be found in Supplementary Table~\ref{tab:DeviceParameters}).
    As before the S\textsubscript{21} measurements (middle column) show the transmission associated with ccw phase-matched directionality within the resonator. We see that as the acousto-optic scattering rate $G_{ph}$ increases, there is further a splitting of the mode induced in the ccw direction, leading to an internal chiral density of states. As a consequence, we simultaneously observe that in the cw direction (S\textsubscript{12} measurements in first column) the optical mode recovers significantly from the undesirable doublet state back to a singlet Lorentzian shape -- even though we did not perform any action to influence the mode in this direction -- and additionally the reflection (third column) from the resonator is eliminated. Both pieces of evidence point to a suppression of Rayleigh backscattering within the resonator.
    }
    \label{fig:4}
    \end{adjustwidth*}
\end{figure}
We now explicitly show the reduction in backscattered light and near-complete mode recovery within a device having a much larger rate of disorder-induced backscattering.
To do so, we also modify our experimental setup by adding a circulator between the laser and the WGR, i.e. before Port 1, to enable simultaneous measurement of light back-reflected from within the resonator (Supplementary Fig.~\ref{supfig:3})
Experimental data are presented in Fig.~\ref{fig:4}, showing forward transmission, reverse transmission, and back-reflection measurements through the waveguide. Only a single reflection measurement is necessary since the reflected signals for the forward and backward propagating directions are the same (see Supplement \S\ref{sec:Evolution}).
We focus here on the TE\textsubscript{00} mode as its linewidth is comparable with the backscattering rate, and it presents a clear visual of the undesirable mode splitting. 
The experiment was performed with a TE\textsubscript{00}/TE\textsubscript{10} mode pair with initial frequency separation of 3.2 GHz. The RF input to the transducer is set at 3 GHz since that was experimentally determined to be the frequency that produces the largest acousto-optic coupling in this device.
In spite of this frequency mismatch (the robustness to frequency mismatches is discussed in \cite{Sohn_Orsel_Bahl_2021}) we do observe a significant phonon-induced hybridization of the forward optical states. This can be discerned from the further splitting of the doublet in the cw phase-matched direction.
From fitting the experimental data (all numerical quantities involved with our experiments are in Supplementary Table~\ref{tab:DeviceParameters}) we estimate that the intrinsic linewidth of the TE\textsubscript{00} mode is $\approx 0.12$ GHz, and the backscattering rates are $V_1 \approx 0.17$ GHz and $V_2 \approx 0.27$ GHz  
following the previous convention in Figs.~\ref{fig:2} and \ref{fig:3}. 
The highest RF drive level tested was up to 27 dBm which we estimate corresponds to a $G_{ph} \approx 0.87$ GHz, and is limited by the power handling capability of the IDT.
At this point $G_{ph} \gg V_1, V_2$ and the TE\textsubscript{00} mode is seen to recover to a Lorentzian with $\approx 0.15$ GHz linewidth, which is an almost complete recovery to the intrinsic loss rate of 0.12 GHz.
Most importantly, we note a significant reduction in the back-reflected light returned to port 1, which is a direct observation confirming near-complete suppression of  disorder-induced Rayleigh backscattering is suppressed within the WGR.

\vspace{12pt}

Disorder-induced backscattering is a considerable technical challenge for integrated photonics platforms. Our experiments demonstrate that the breaking of TRS within integrated wave-guiding structures -- here, a resonator -- can suppress this unwanted scattering effect even when disorder and defects may be present. 
This is an extremely important capability since isolators and circulators can only discard back-reflected photons and do nothing to suppress the scattering from taking place at its point of origination.
While the technique we show does require a sacrifice of the modes in a counter-propagating direction, there is still tremendous potential for systems where unidirectional light transport is necessary.
For instance, the application of such a technique on a non-resonant waveguide could ensure very high fidelity unidirectional light transport from an integrated laser to the rest of the optical system, without requiring an isolator. Backscattering free inter-chip light transport might also be accomplished by similarly breaking TRS across a butt coupled waveguide-waveguide interface.
Finally, we note that our acousto-optic approach in lithium niobate is by no means exclusive, and there are plenty of TRS-breaking alternatives that could offer good solutions to suppressing disorder-induced backscattering in integrated photonics. These include optomechanics~\cite{Kim:2015,Kim_2017}, 
electro-optic interaction~\cite{lira2012,Sounas:14}, 
and on-chip magneto-optic devices~\cite{Ross:11, Ghosh:12aa, Huang:17, Zhang:2017wq}.

\vspace{24pt}

{\footnotesize \putbib}

{\footnotesize \bibliography{thesisrefsol}}
\end{bibunit}
\section*{Acknowledgments}

This work was sponsored by the Defense Advanced Research Projects Agency (DARPA) grant FA8650-19-2-7924 and the Air Force Office of Scientific Research (AFOSR) grant FA9550-19-1-0256. GB would additionally like to acknowledge support from the Office of Naval Research (ONR) Director for Research Early Career grant N00014-17-1-2209 and the Presidential Early Career Award for Scientists and Engineers. We also thank Josephine Melia for assistance with preliminary experiments on this topic.

\section*{Data availability}

The data that support the findings of this study are available from the corresponding author upon reasonable request.

\section*{Author contributions}

O.E.Ö. and G.B. jointly conceived the concept, O.E.Ö. and J.N. performed the theoretical analysis. O.E.Ö. performed the device fabrication and conducted the experimental measurements. O.E.Ö. and J.N. analysed the data. All the authors contributed to writing the paper. G.B. supervised all aspects of this project.

\FloatBarrier

\newpage

\renewcommand*{\citenumfont}[1]{S#1}
\renewcommand*{\bibnumfmt}[1]{[S#1]}
\newcommand{\beginsupplement}{%
        \setcounter{table}{0}
        \renewcommand{\thetable}{S\arabic{table}}%
        \setcounter{figure}{0}
        \renewcommand{\thefigure}{S\arabic{figure}}%
        \setcounter{equation}{0}
        \renewcommand{\theequation}{S\arabic{equation}}
        \setcounter{section}{0}
        \renewcommand{\thesection}{S\arabic{section}}%
}

\beginsupplement
\begin{bibunit}

\begin{center}

\Large{\textbf{Supplementary Information: \\  Electrically-Controlled Suppression \\ of Rayleigh Backscattering \\ in an Integrated Photonic Circuit}} \\
    
\vspace{24pt}

\large{
{Oğulcan E. Örsel $^1$}, {Jiho Noh $^2$},
and {Gaurav Bahl $^2$}} \\
\vspace{12pt}
    \footnotesize{$^1$ Department of Electrical $\&$ Computer Engineering,} \\
    \footnotesize{$^2$ Department of Mechanical Science and Engineering,} \\
    \footnotesize{University of Illinois at Urbana–Champaign, Urbana, IL 61801 USA,} \\
\end{center}

\vspace{24pt}

\section{Two-level photonic system with acousto-optic and Rayleigh interaction }
\label{sec:Modelling}

\vspace{12pt}

Our photonic system consists of two optical modes, TE\textsubscript{00} ($\omega_2$, $k_2$) and TE\textsubscript{10} ($\omega_1$, $k_1$) that are supported within a racetrack resonator. These modes are coupled with an acoustic pump ($\Omega$, $q$) that bridges the frequency ($\Omega=\omega_1-\omega_2$) and momentum ($q=k_2-k_1$) gap between them. Since this coupling (or the three-wave mixing process) is satisfied in a unidirectional manner, we use the term ``Forward" for the phase-matched direction and ``Backward" for the non-phase-matched direction. For the forward direction with $\Omega=\omega_1-\omega_2$, we can then write the interaction Hamiltonian for the acousto-optic interaction as \cite{Agrawal2013,Bochmann_2013,Kim_Taylor_Bahl_2019,Sohn_Orsel_Bahl_2021};

\begin{equation}
H_{in}^{AO}=\hbar\left(\frac{g_{ph}}{2}\hat{a}_{1+}^\dagger\hat{a}_{2+}\hat{b}+\frac{g_{ph}^*}{2}\hat{a}_{1+}\hat{a}_{2+}^{\dagger}\hat{b}^\dagger\right)
\end{equation}

We define the single photon acousto-optic coupling rate between the modes as $g\textsubscript{ph} \propto \delta(k_2-k_1-q)\int
\mathcal{E}\textsubscript{1}(\textbf{r}_\bot)\mathcal{E}\textsubscript{2}(\textbf{r}_\bot)u(\textbf{r}_\bot)d\boldsymbol{r}_\bot$~\cite{Agrawal2013}, where $\mathcal{E}\textsubscript{1}(\boldsymbol{r}_\bot)$, $\mathcal{E}\textsubscript{2}(\boldsymbol{r}_\bot)$ and $u(\boldsymbol{r}_\bot)$ are the transverse mode profiles of the optical and acoustic modes, respectively. 
Here $\delta(k_2-k_1-q)$ represents the unidirectional phase matching feature of the process, $\hat{a}_{1+}^\dagger$($\hat{a}_{1+}$), $\hat{a}_{2+}^\dagger$($\hat{a}_{2+}$) and $\hat{b}^\dagger$($\hat{b}$) represent creation (annihilation) operators for TE\textsubscript{10}, TE\textsubscript{00} photons and acoustic phonons respectively. In addition to the acousto-optic coupling rate, we consider the interaction of the forward optical modes, TE\textsubscript{00+} ($\hat{a}_{2+}$) and TE\textsubscript{10+} ($\hat{a}_{1+}$), with their time-reversal counterparts, TE\textsubscript{00-} ($\hat{a}_{2-}$) and TE\textsubscript{10-} ($\hat{a}_{1-}$), as well. The coupling is induced due to surface roughness or internal inhomogeneities and is usually referred to as Rayleigh scattering. Under the dipole approximation, we can write the interaction Hamiltonian for this case as,

\begin{equation}
H_{in}^{R}=\hbar\frac{V_1}{2}\left(\hat{a}^\dagger_{1+}\hat{a}_{1-}+\hat{a}^\dagger_{1-}\hat{a}_{1+}\right)+
\hbar\frac{V_2}{2}\left(\hat{a}^\dagger_{2+}\hat{a}_{2-}+\hat{a}^\dagger_{2-}\hat{a}_{2+}\right)
\end{equation}

Here we define V\textsubscript{1} and V\textsubscript{2} as the backscattering rates for the TE\textsubscript{10} and TE\textsubscript{00} modes, respectively. These coupling rates depend on the effective mode volume, the overlap of the scatterer with the optical mode, and the optical frequency. For our case, the optical modes are expected to show different backscattering rates, and this is verified by the experiments.

\vspace{12pt}

Having defined the interaction Hamiltonian for both cases, we can now write the Heisenberg-Langevin equations of our system. While doing so, we also treat the Heisenberg-Langevin equations classically by making substitutions as follows: $\hat{a}_{1+}$($\hat{a}_{1-}$)~$\rightarrow$~$a_{1+}$($a_{1-}$), $\hat{a}_{2+}$($\hat{a}_{2-}$)~$\rightarrow$~$a_{2+}$($a_{2-}$) and $\hat{b}\, g_{ph}/{2}$~$\rightarrow$~${b \, g_{ph}}/{2}$. Here, $a_{1+}$($a_{1-}$) and $a_{2+}$($a_{2-}$) are intracavity field amplitudes of the forward (backward) TE\textsubscript{10} and TE\textsubscript{00} modes respectively. Also, $b$ represents the steady-state amplitude of the acoustic wave under non-depleted RF pump approximation. Furthermore, we also combine $b$ with the single photon acousto-optic coupling rate and re-write it as $G_{ph}/2$. Then, the equations of motion become:

\begin{align}
\hspace{-20pt}
\partial_{t}
\begin{pmatrix}
a_1^+ \\
a_2^+ \\
a_1^- \\
a_2^- \\
\end{pmatrix}
=&
\left(\begin{matrix}
-\frac{\kappa_{1}}{2} -i\omega_1 & -i\frac{G_{ph}}{2}e^{-i\Omega t}-\Gamma_c\\
-i\frac{G_{ph}}{2}e^{i\Omega t}-\Gamma_{c} & -\frac{\kappa_{2}}{2} -i\omega_2  \\
-i\frac{V_{1}}{2} & 0  \\
 0 & -i\frac{V_{2}}{2} \\
\end{matrix} 
\begin{matrix}
-i\frac{V_{1}}{2} & 0 \\
0 & -i\frac{V_{2}}{2} \\
-\frac{\kappa_{1}}{2} -i\omega_1 & -i\frac{G_{ph}}{2}e^{i\Omega t}-\Gamma_c \\
-i\frac{G_{ph}}{2}e^{-i\Omega t}-\Gamma_c & -\frac{\kappa_{2}}{2} -i\omega_2 \\
\end{matrix}\right)
\begin{pmatrix}
a_1^+ \\
a_2^+ \\
a_1^- \\
a_2^- \\
\end{pmatrix} \nonumber\\
+
&\begin{pmatrix}
\sqrt{\kappa_{ex1}}s_{in}^+ \\
\sqrt{\kappa_{ex2}}s_{in}^+ \\
\sqrt{\kappa_{ex1}}s_{in}^- \\
\sqrt{\kappa_{ex2}}s_{in}^-
\label{eq:general_equation}
\end{pmatrix}
\end{align}

Here $\kappa_1$ ($\kappa_2$) and $\kappa_{ex1}$ ($\kappa_{ex2}$) represent the total loss rate and the external coupling rate of the TE\textsubscript{10} (TE\textsubscript{00}) modes. $\omega_1$ ($\omega_2$) is the resonant frequency of TE\textsubscript{10} (TE\textsubscript{00}) mode, and $G\textsubscript{ph}$ is the phonon enhanced acousto-optic coupling rate. We also added $\Gamma_c=\sqrt{\kappa_{ex1}\kappa_{ex2}}/2$~\cite{Zhang_Darmawan_Mei_Zhang_2010}, which describes the coupling between TE\textsubscript{00} and TE\textsubscript{10} modes due to the presence of the probe waveguide.

\vspace{12pt}

To solve the above equation \ref{eq:general_equation}, we need to take the Fourier transform of both sides. Since our system is a linear time-varying system, we first need to decompose the optical fields into Fourier components. For that purpose, we consider the TE\textsubscript{10} mode to be located at a higher frequency than the TE\textsubscript{00} mode and expand the Fourier components as,

\begin{equation}
a_1^+=a_{1,0}^+e^{-i\omega_l t}+a_{1,+1}^+e^{-i(\omega_l+\Omega)t}
\end{equation}
\begin{equation}
a_2^+=a_{2,0}^+e^{-i\omega_l t}+a_{2,-1}^+e^{-i(\omega_l-\Omega)t}
\end{equation}
\begin{equation}
a_1^-=a_{1,0}^-e^{-i\omega_l t}+a_{1,-1}^-e^{-i(\omega_l-\Omega)t}
\end{equation}

\begin{equation}
a_2^-=a_{2,0}^-e^{-i\omega_l t}+a_{2,+1}^-e^{-i(\omega_l+\Omega)t}
\end{equation}

We then substitute these equations into equation \ref{eq:general_equation} to obtain three sets of matrix equations at different frequencies. At the optical carrier frequency $\omega_l$, we have :
\begin{multline}
\hspace{-20pt}
\partial_{t}
\begin{pmatrix}
a_{1,0}^+e^{-i\omega_l t} \\
a_{2,0}^+e^{-i\omega_l t} \\
a_{1,0}^-e^{-i\omega_l t} \\
a_{2,0}^-e^{-i\omega_l t} \\
\end{pmatrix}
=
\begin{pmatrix}
-\frac{\kappa_1}{2} -i\omega_1 & -\Gamma_c & -i\frac{V_1}{2} & 0 \\
-\Gamma_c & -\frac{\kappa_2}{2} -i\omega_2 & 0 & -i\frac{V_2}{2} \\
-i\frac{V_1}{2} & 0 & -\frac{\kappa_1}{2} -i\omega_1 & -\Gamma_c \\
 0 & -i\frac{V_2}{2} & -\Gamma_c & -\frac{\kappa_2}{2} -i\omega_2\\
\end{pmatrix}
\begin{pmatrix}
a_{1,0}^+e^{-i\omega_l t} \\
a_{2,0}^+e^{-i\omega_l t} \\
a_{1,0}^-e^{-i\omega_l t} \\
a_{2,0}^-e^{-i\omega_l t} \\
\end{pmatrix}
\\
+
\begin{pmatrix}
0 & -i\frac{G_{ph}}{2} & 0 & 0 \\
-i\frac{G_{ph}}{2} & 0 & 0 & 0\\
0 & 0 & 0 & -i\frac{G_{ph}}{2} \\
 0 & 0 & -i\frac{G_{ph}}{2} & 0\\
\end{pmatrix}
\begin{pmatrix}
a_{1,+1}^+e^{-i\omega_l t} \\
a_{2,-1}^+e^{-i\omega_l t} \\
a_{1,-1}^-e^{-i\omega_l t} \\
a_{2,+1}^-e^{-i\omega_l t} \\
\end{pmatrix}
+
\begin{pmatrix}
\sqrt{\kappa_{ex1}}s_{in}^+e^{-i\omega_l t} \\
\sqrt{\kappa_{ex2}}s_{in}^+e^{-i\omega_l t} \\
\sqrt{\kappa_{ex1}}s_{in}^-e^{-i\omega_l t} \\
\sqrt{\kappa_{ex2}}s_{in}^-e^{-i\omega_l t} \\
\end{pmatrix}
\end{multline}
Similarly, at $\omega_l - \Omega$:
\begin{multline}
\hspace{-20pt}
\partial_{t}
\begin{pmatrix}
0 \\
a_{2,-1}^+e^{-i(\omega_l-\Omega) t} \\
a_{1,-1}^-e^{-i(\omega_l-\Omega) t} \\
0 \\
\end{pmatrix}
=
\begin{pmatrix}
-\frac{\kappa_1}{2} -i\omega_1 & -\Gamma_c & -i\frac{V_1}{2} & 0 \\
-\Gamma_c & -\frac{\kappa_2}{2} -i\omega_2 & 0 & -i\frac{V_2}{2} \\
-i\frac{V_1}{2} & 0 & -\frac{\kappa_1}{2} -i\omega_1 & -\Gamma_c \\
 0 & -i\frac{V_2}{2} & -\Gamma_c & -\frac{\kappa_2}{2} -i\omega_2\\
\end{pmatrix}\\
\times\begin{pmatrix}
0 \\
a_{2,-1}^+e^{-i(\omega_l-\Omega) t} \\
a_{1,-1}^-e^{-i(\omega_l-\Omega) t} \\
0 \\
\end{pmatrix}
+
\begin{pmatrix}
0 & -i\frac{G_{ph}}{2} & 0 & 0 \\
-i\frac{G_{ph}}{2} & 0 & 0 & 0\\
0 & 0 & 0 & -i\frac{G_{ph}}{2} \\
 0 & 0 & -i\frac{G_{ph}}{2} & 0\\
\end{pmatrix}
\begin{pmatrix}
a_{1,0}^+e^{-i(\omega_l -\Omega)t} \\
0 \\
0 \\
a_{2,0}^-e^{-i(\omega_l-\Omega )t} \\
\end{pmatrix}
\end{multline}
Similarly, we have at $\omega_l + \Omega$:
\begin{multline}
\hspace{-20pt}
\partial_{t}
\begin{pmatrix}
a_{1,+1}^+e^{-i(\omega_l+\Omega) t} \\
0 \\
0 \\
a_{2,+1}^-e^{-i(\omega_l+\Omega) t} \\
\end{pmatrix}
=
\begin{pmatrix}
-\frac{\kappa_1}{2} -i\omega_1 & -\Gamma_c & -i\frac{V_1}{2} & 0 \\
-\Gamma_c & -\frac{\kappa_2}{2} -i\omega_2 & 0 & -i\frac{V_2}{2} \\
-i\frac{V_1}{2} & 0 & -\frac{\kappa_1}{2} -i\omega_1 & -\Gamma_c \\
 0 & -i\frac{V_2}{2} & -\Gamma_c & -\frac{\kappa_2}{2} -i\omega_2\\
\end{pmatrix}\\
\times\begin{pmatrix}
a_{1,+1}^+e^{-i(\omega_l+\Omega) t} \\
0\\
0 \\
a_{2,+1}^-e^{-i(\omega_l+\Omega)t} \\
\end{pmatrix}
+
\begin{pmatrix}
0 & -i\frac{G_{ph}}{2} & 0 & 0 \\
-i\frac{G_{ph}}{2} & 0 & 0 & 0\\
0 & 0 & 0 & -i\frac{G_{ph}}{2} \\
 0 & 0 & -i\frac{G_{ph}}{2} & 0\\
\end{pmatrix}
\begin{pmatrix}
0 \\
a_{2,0}^+e^{-i(\omega_l+\Omega) t}\\
a_{1,0}^-e^{-i(\omega_l+\Omega) t} \\
0 \\
\end{pmatrix}
\end{multline}
These 12 equations describe the state of our two-level photonic system under both acousto-optic and Rayleigh scattering. From these equations, we can derive the intracavity field vector at the carrier frequency ($\omega_l$) as:

\begin{centering}
\begin{equation}
S=\begin{pmatrix}
-\frac{\kappa_1}{2} -i\omega_1 & -\Gamma_c & -i\frac{V_1}{2} & 0 \\
-\Gamma_c & -\frac{\kappa_2}{2} -i\omega_2 & 0 & -i\frac{V_2}{2} \\
-i\frac{V_1}{2} & 0 & -\frac{\kappa_1}{2} -i\omega_1 & -\Gamma_c \\
 0 & -i\frac{V_2}{2} & -\Gamma_c & -\frac{\kappa_2}{2} -i\omega_2\\
\end{pmatrix}
\end{equation}
\end{centering}
\begin{centering}
\begin{equation}
C_n=\begin{pmatrix}
0 & 0 & 0 & 0 \\
-i\frac{G_{ph}}{2} & 0 & 0 & 0\\
0 & 0 & 0 & -i\frac{G_{ph}}{2} \\
 0 & 0 & 0 & 0\\
\end{pmatrix},
C_p=\begin{pmatrix}
0 & -i\frac{G_{ph}}{2} & 0 & 0 \\
0 & 0 & 0 & 0\\
0 & 0 & 0 & 0 \\
 0 & 0 & -i\frac{G_{ph}}{2} & 0\\
\end{pmatrix}
\end{equation}
\end{centering}
\begin{centering}
\begin{equation}
\hspace{-60pt}
\gamma_n
=
\left[-(\omega_l-\Omega)\mathbb{1}-S\right]^{-1}C_n
\hspace{10pt},\hspace{10pt}
\gamma_p
=
\left[-(\omega_l+\Omega)\mathbb{1}-S\right]^{-1}C_p
\end{equation}
\end{centering}
\begin{centering}
\begin{equation}
\hspace{-60pt}
A_0
=
\left[-\omega_l\mathbb{1}-S-C_p\gamma_n-C_n\gamma_p\right]^{-1}
\begin{pmatrix}
\sqrt{\kappa_{ex1}}s_{in}^+ \\
\sqrt{\kappa_{ex2}}s_{in}^+ \\
\sqrt{\kappa_{ex1}}s_{in}^- \\
\sqrt{\kappa_{ex2}}s_{in}^- \\
\end{pmatrix}
\end{equation}
\end{centering}
We then calculate the output spectrum at the carrier frequency as:

\begin{centering}
\begin{equation}
\hspace{-60pt}
s_{out}^+
=
s_{in}^+-
\begin{pmatrix}
\sqrt{\kappa_{ex1}} \\
\sqrt{\kappa_{ex2}} \\
0\\
0\\
\end{pmatrix}
^T
A_0
\end{equation}
\end{centering}
\begin{centering}
\begin{equation}
\hspace{-60pt}
r_{out}^+
=
-
\begin{pmatrix}
0 \\
0 \\
\sqrt{\kappa_{ex1}} \\
\sqrt{\kappa_{ex2}} \\
\end{pmatrix}
^T
A_0
\end{equation}
\end{centering}
\begin{centering}
\begin{equation}
\hspace{-60pt}
s_{out}^-
=
s_{in}^--
\begin{pmatrix}
0 \\
0 \\
\sqrt{\kappa_{ex1}} \\
\sqrt{\kappa_{ex2}} \\
\end{pmatrix}
^T
A_0
\end{equation}
\end{centering}
\begin{centering}
\begin{equation}
\hspace{-60pt}
r_{out}^-
=
-
\begin{pmatrix}
\sqrt{\kappa_{ex1}} \\
\sqrt{\kappa_{ex2}} \\
0\\
0\\
\end{pmatrix}
^T
A_0
\end{equation}
\end{centering}
Here, we define $s_{in}^{+} (s_{in}^{-})$, $s_{out}^{+} (s_{out}^{-})$ and $r_{out}^{+} (r_{out}^{-})$ for a two port system (Port 1 and Port 2.). For an excitation at port 1 (port 2) with $s_{in}^{+} (s_{in}^{-})$ , $s_{out}^{+} (s_{out}^{-})$ is the transmitted light at port 2 (port 1), and $r_{out}^{+} (r_{out}^{-})$ is the reflected light at port 1 (port 2).

\vspace{12pt}

For a better understanding of our model, we consider the experimental results given in Fig.~\ref{fig:3}. To simplify the analysis, we assume that the external coupling rate of the TE\textsubscript{00} mode is approximately zero ($\kappa_{ext2}\approx0$), and the modal loss rates ($\kappa_1$ and $\kappa_2$) are similar to each other with $V_1$ being ignored.
Also, we are interested in the optical output at the input (carrier) frequency $s_{out,0}$  (ignoring the sidebands $s_{out,1}$) for the phase-matched direction. We can then write:
\begin{align}
    \label{eqn:outputeqn}
    s_{out,0}^+ = s_{in}^+ - \sqrt{\kappa_{ex1}} \, a_{1,0,+}
\end{align}
where the TE$_{10}$ intracavity field at the input frequency ($a_{1,0+}$), and it can be expressed as:
\begin{align}
    \label{eqn:a2_phasematched}
    a_{1,0+} = 
        \frac
            {\sqrt{\kappa_{ex1}} \, \left[ \cfrac{V_2^2}{4}+\left(\cfrac{\kappa_1}{2} - i\Delta\right)^2 \right]}
            {\left[\cfrac{G_{ph}^2}{4}+\cfrac{V_2^2}{4}+\left(\cfrac{\kappa_1}{2} - i\Delta\right)^2 \right]\left(\cfrac{\kappa_1}{2} - i\Delta \right)} 
        \, s_{in}
\end{align}
Here, $\Delta = \omega_l - \omega_1$ is the optical detuning from the TE$_{10}$ mode. We can then simplify the TE$_{10}$ intracavity field to three diagonalized intracavity fields by assuming $G_{ph}\gg V_2$;

\begin{equation}
  \begin{gathered}[b]
    a_{1,0+} = 
        \frac{V_2^2}{G_{ph}^2}\left(\frac{\sqrt{\kappa_{ex2}}}{\nicefrac{\kappa_1}{2} - i\Delta}\right) + \frac{G_{ph}^2-V_2^2}{G_{ph}^2}\left[\frac{\sqrt{\kappa_{ex2}}/2}{\nicefrac{\kappa_1}{2} - i \left(\Delta+ \nicefrac{G_{ph}}{2} \right)}\right] \\
        + \frac{G_{ph}^2-V_2^2}{G_{ph}^2}\left[\frac{\sqrt{\kappa_{ex2}}/2}{\nicefrac{\kappa_1}{2} - i \left(\Delta- \nicefrac{G_{ph}}{2} \right)}\right]s_{in}
        \label{eq:three_mode}
  \end{gathered}
\end{equation}
Here we can easily see that the cavity susceptibility is now modified from a single Lorentzian response and instead exhibits three distinctive Lorentzian responses. If the acousto-optic coupling rate is large enough ($G_{ph} \gg \sqrt{\kappa_1\kappa_2}$ and $G_{ph} \gg V_2$ ), these responses correspond to the dressed states of TE\textsubscript{10} mode and the backscattered TE\textsubscript{00} mode, which coincides with the original TE\textsubscript{10} in the forward direction. As apparent from equation \ref{eq:three_mode}, the cavity susceptibilities for the dressed states increase and split further in frequency when the $G_{ph}$ increases. On the other hand, the cavity susceptibility of the backward TE\textsubscript{00} mode decreases, demonstrating the suppression of the backscattering. In other words, the optical states for the forward modes change. This modification produces a reduction of the spectral overlap between the counter-propagating modes.
As discussed in the main text (Fig.~\ref{fig:3}b), the optical transmission measurement verifies this behavior as a reduced height for the central dip and increased splitting for the outer modes (i.e. the dressed states).

\section{Example evolution of transmission and reflection coefficients}
\label{sec:Evolution}

\begin{figure}[b!]
    \begin{adjustwidth*}{-1in}{-1in}
    \hsize=\linewidth
    \includegraphics[width=1.2\textwidth, clip=true, trim=0.6in 0.2in 0.2in 0.2in]{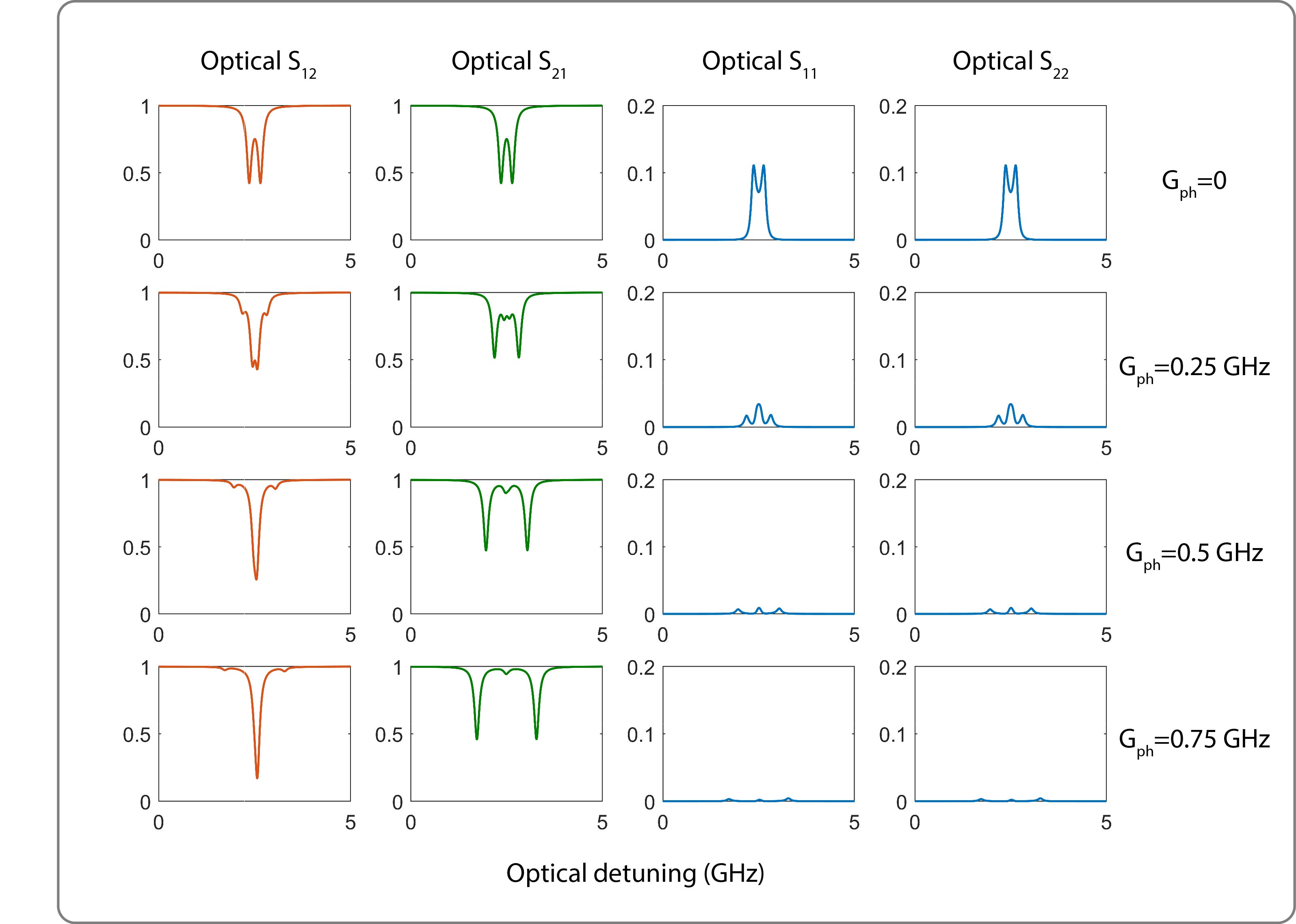} 
    \centering
    \caption{
    \textbf{Example prediction of backscattering suppression and observation of optical S-parameters (reflection and transmission coefficients)}
     Selected parameters are $\kappa_1=\kappa_2= 0.1$ GHz, $V_1/2=V_2/2=0.15$ GHz, $\kappa_{ex1}=\kappa_{ex2}=0.05$ GHz, $\omega_1-\omega_2=5$ GHz and $\Omega=5$ GHz. We show the evolution of the optical S-parameters under increasing acousto-optic coupling rate $G_{ph}$.
    }
    \label{supfig:1}
    \end{adjustwidth*}
\end{figure}

Here, we consider an example case to understand the system dynamics under increasing acousto-optic coupling rate. For optical modes, we assume they have matched optical parameters: $\kappa_1=\kappa_2= 0.1$ GHz, $V_1/2=V_2/2=0.15$ GHz, $\kappa_{ex1}=\kappa_{ex2}=0.05$ GHz,  $\omega_1-\omega_2=5$ GHz and $\Omega=5$ GHz. We plot only one of the modes since they have identical optical parameters. In Fig.~\ref{supfig:1}, we present the optical S parameters of the system for increasing acousto-optic coupling rate ($G\textsubscript{ph}$). 
For $G\textsubscript{ph}=0$, the modes exhibit their intrinsic backscattering induced mode splitting, which is clearly resolved since the backscattering rate is larger than the optical loss rates. As we increase $G\textsubscript{ph}$, we see that the mode in the non-phase matched direction recovers from this undesirable doublet state to a singlet state, just as is observed in our experiments. The mode in the phase-matched direction, i.e. where the $G_{ph}$ is active, splits even further due to strong acousto-optic mode hybridization. The unwanted reflections due to intrinsic Rayleigh scattering also monotonically decrease. One crucial observation is that the optical reflection coefficients are the same (optical S\textsubscript{11}=S\textsubscript{22}) for opposing directions due to the symmetry of the reflection process.

\section{Transformation of the frequency basis for equivalent system representation}
\label{sec:DerivationHamiltonian}

\begin{figure}[bh!]
    \begin{adjustwidth*}{-1in}{-1in} 
    \hsize=\linewidth
    \includegraphics[width=1.1\textwidth]{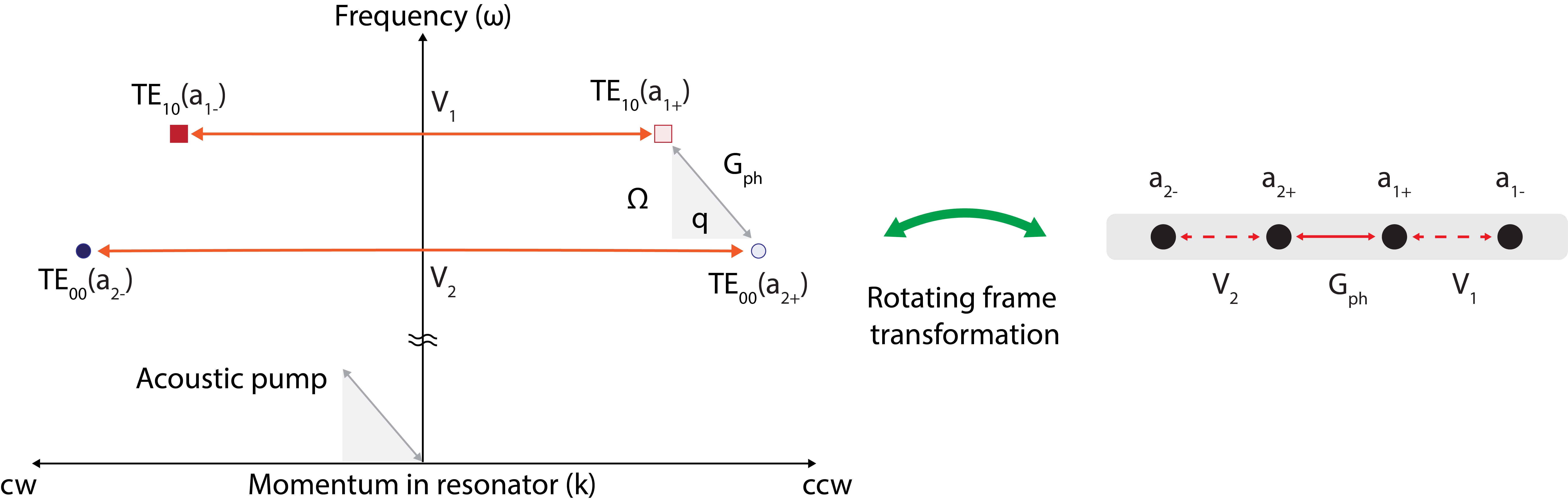}
    \centering
    \caption{
    \textbf{Transformation of a two-level photonic system to a 1D coupled resonator chain.}
    The forward and backward optical modes of a two-level photonic system have distinct frequencies and momentum, and they cannot be represented with a static Hamiltonian (i.e., without oscillating terms). However, with appropriate choice of a rotating frame as $e^{i\Omega t}$, the two-level photonic system can be described by a 1D coupled resonator chain.
    }
    \label{supfig:transformation}
    \end{adjustwidth*}
\end{figure}

The two-level photonic system with backscattering can be transformed into a coupled resonator chain with 4-sites. To show this equivalence, we start with the equations of motions in the original static frame where we neglect any acousto-optic interaction in the non-phase matched direction:

\begin{equation}
    \frac{d}{dt}
    \begin{pmatrix}
        a_{1+}\\ a_{2+}\\a_{1-}\\a_{2-}
    \end{pmatrix}
    = -i
    \begin{pmatrix}
        \omega_{1}-i\frac{\kappa_{1}}{2} & \frac{G_{ph}}{2}e^{-i\Omega t} & \frac{V_{1}}{2} & 0 \\
     \frac{G_{ph}}{2} e^{i\Omega t} & \omega_{2}-i\frac{\kappa_{2}}{2} & 0 & \frac{V_{2}}{2} \\
     \frac{V_{1}}{2} & 0 & \omega_{1}-i\frac{\kappa_{1}}{2} & 0 \\
     0 & \frac{V_{2}}{2} & 0 &\omega_{2}-i\frac{\kappa_{2}}{2}
    \end{pmatrix}
    \begin{pmatrix}
        a_{1+}\\ a_{2+}\\a_{1-}\\a_{2-}
    \end{pmatrix}.
    \label{eq:StaticFrameEoM}
\end{equation}
Then, we change the frame of reference, where we choose a frame in which modes are transformed as $a_{1\pm} \rightarrow a_{1\pm}$ and $a_{2\pm}\rightarrow a_{2\pm}e^{i\Omega t}$.
As a consequence, each equation of motion transforms as:
\begin{align}
\dot{a_{1+}} &= -i\left[ \left( \omega_{1}-i\frac{\kappa_{1}}{2}\right)a_{1+} + \frac{G_{ph}}{2}\cancel{e^{-i\Omega t}} (a_{2+}\cancel{e^{i\Omega t}}) + \frac{V_{1}}{2}a_{1-} \right] \notag  \\
&= -i\left[ \left( \omega_{1}-i\frac{\kappa_{1}}{2}\right)a_{1+} + \frac{G_{ph}}{2} a_{2+} + \frac{V_{1}}{2}a_{1-} \right] \\
\frac{d}{dt}\left(a_{2+}e^{i\Omega t}\right) & =\dot{a_{2+}}\cancel{e^{i\Omega t}}+i\Omega a_{2+}\cancel{e^{i\Omega t}} \notag  \\
& =-i\left[ \frac{G_{ph}}{2} \cancel{e^{i\Omega t}}a_{1+}+\left(\omega_{2}-i\frac{\kappa_{2}}{2} \right)a_{2+}\cancel{e^{i\Omega t}} + \frac{V_{2}}{2}a_{2-}\cancel{e^{i\Omega t}}\right] \notag  \\
\therefore \dot{a_{2+}} & = -i\left[ \frac{G_{ph}}{2} a_{1+} + \left( \omega_{2}+\Omega-i\frac{\kappa_{2}}{2}\right)a_{2+}+\frac{V_{2}}{2}a_{2-}\right] \\
\dot{a_{1-}} & =-i\left[\frac{V_{1}}{2}a_{1+}+\left( \omega_{1}-i\frac{\kappa_{1}}{2} \right)a_{1-} \right]\\
\frac{d}{dt}\left(a_{2-}e^{i\Omega t}\right)&=\dot{a_{2-}}\cancel{e^{i\Omega t}}+i\Omega a_{2-}\cancel{e^{i\Omega t}} \notag  \\
&=-i\left[\frac{V_{2}}{2}a_{2+}\cancel{e^{i\Omega t}}+\left(\omega_{2}-i\frac{\kappa_{2}}{2} \right)a_{2-}\cancel{e^{i\Omega t}}\right] \notag  \\
\therefore a_{2-} & = -i\left[\frac{V_{2}}{2}a_{2+} + \left( \omega_{2}+\Omega-i\frac{\kappa_{2}}{2}\right)a_{2-}\right]
\end{align}
Therefore, in the chosen frame of reference, the equation of motion becomes:
\begin{equation}
    \frac{d}{dt}
    \begin{pmatrix}
        a_{1+}\\ a_{2+}\\a_{1-}\\a_{2-}
    \end{pmatrix}
    = -i
    \begin{pmatrix}
        \omega_{1}-i\frac{\kappa_{1}}{2} & \frac{G_{ph}}{2} & \frac{V_{1}}{2} & 0 \\
     \frac{G_{ph}}{2}  & \omega_{2}+\Omega - i\frac{\kappa_{2}}{2} & 0 & \frac{V_{2}}{2} \\
     \frac{V_{1}}{2} & 0 & \omega_{1}-i\frac{\kappa_{1}}{2} & 0 \\
     0 & \frac{V_{2}}{2} & 0 &\omega_{2} + \Omega - i\frac{\kappa_{2}}{2}
    \end{pmatrix}
    \begin{pmatrix}
        a_{1+}\\ a_{2+}\\a_{1-}\\a_{2-}
    \end{pmatrix},
    \label{eq:RotatingFrameEoM}
\end{equation}
where the off-diagonal terms, which describe the interaction between different modes, form the interaction Hamiltonian. From equation~\ref{eq:RotatingFrameEoM}, we see that the equivalence that is described in Fig~\ref{supfig:transformation} holds.

\section{Similarities between the photonic molecule and the SSH model}
\label{sec:SimilaritiesSSH}

\begin{figure}[ht!]
    \begin{adjustwidth}{-1in}{-1in} 
    \hsize=\linewidth
    \centering
    \includegraphics[width=\textwidth]{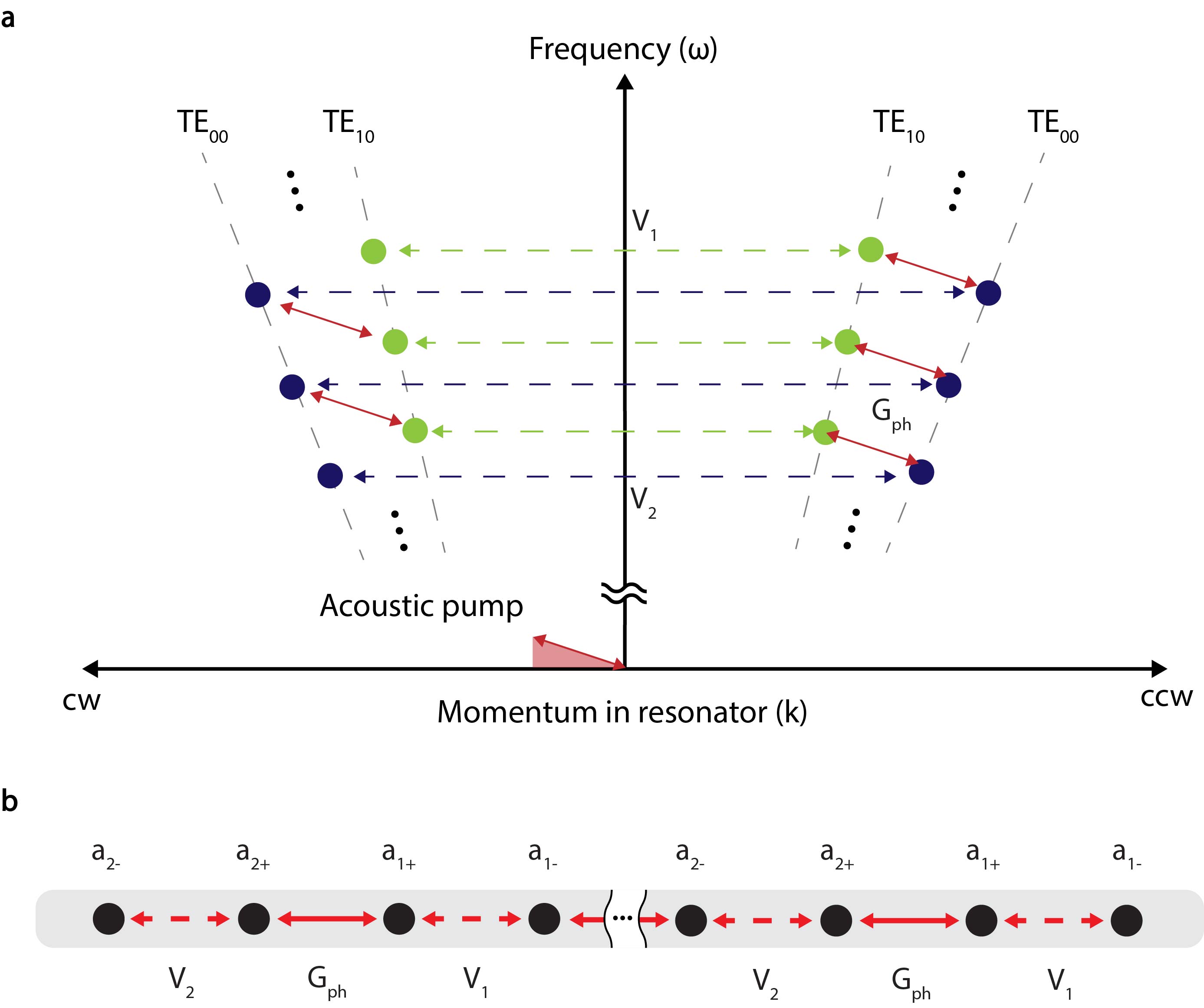}
    \caption{
    \textbf{Extension of the system into longer chains}
    \textbf{(a)} The band diagram of the modified system shows both the forward and backward phase matching to extend the SSH chain.
    \textbf{(b)} Equivalent SSH chain of the system under rotating frame approximation.
    }
    \label{LongerChain}
    \end{adjustwidth}    
\end{figure}

As described in the main text, the suppression of the backscattering can also be understood in terms of the topological phases of the well-known Su-Schrieffer-Heeger (SSH) model, where the interaction Hamiltonian of the system can be interpreted as a short 1D SSH chain in a rotating reference frame (see Fig.~\ref{supfig:transformation}).
In this model, the backscattering rates ($V_{1}$ and $V_{2}$) and the acousto-optical coupling rate ($G_{ph}$) can be interpreted as intra- and inter-unit-cell couplings, respectively, where the control over $G_{ph}$ allows us to control the competition between the coupling mechanisms within the model, and hence the topological phase of the system.

\vspace{12pt}

For example, the system in a weak acousto-optical coupling regime ($G_{ph}<V_{1},\, V_{2}$) is equivalent to the topologically trivial phase of the SSH chain in which we are only able to observe the doublet bulk modes of the SSH chain.
On the other hand, increasing $G_{ph}$ induces a topological transition of the system, and the system in a strong acousto-optical coupling regime ($G_{ph}\gg V_{1},\, V_{2}$) with phase-matching condition becomes equivalent to the topologically non-trivial phase of the SSH chain.
In this regime, the topologically protected edge modes of the SSH model are predominantly occupied by the backward propagating modes.
Since they are spectrally localized away from the bulk modes, i.e., the forward propagating modes, we observe the retrieval of each singlet backward propagating mode in contrast to the doublet forward propagating modes.
Therefore, by changing the relative strength between the coupling mechanisms, we can effectively suppress the backscattering.
This topological protection of the backward propagating modes breaks as the phase matching condition breaks or as the acousto-optical coupling becomes weaker, which is manifested by the doublet backward propagating modes.
The mechanism of how the topological protection of the backward propagating modes breaks in two cases differ: in the former case, where $\Omega\neq\omega_{1}-\omega_{2}$ the chiral symmetry of the system breaks, and in the latter case, the system enters a topologically trivial phase.

\vspace{12pt}

In this paper, our equivalent SSH chain is composed of 2 unit cells.
However, this chain can be easily extended by considering a larger racetrack resonator, where multiple mode pairs can be coupled in both forward and backward directions (Fig~\ref{LongerChain}).
In a larger racetrack resonator, the FSR of each mode pair (i.e., FSR\textsubscript{1}, FSR\textsubscript{2} $\approx \Omega$) decreases and hence makes it possible to couple forward and backward mode pairs [i.e., $a\textsubscript{1+}(\omega_0+\Omega)$, $a\textsubscript{2+}(\omega_0)$, $a\textsubscript{1-}(\omega_0+\Omega)$, $a\textsubscript{2-}(\omega_0)$] with another one that is separated by one FSR [i.e., $a\textsubscript{1+}(\omega_0+\Omega+\textrm{FSR\textsubscript{1}})$, $a\textsubscript{2+}(\omega_0+\textrm{FSR\textsubscript{2}})$, $a\textsubscript{1-}(\omega_0+\Omega+\textrm{FSR\textsubscript{1}})$, $a\textsubscript{2-}(\omega_0+\textrm{FSR\textsubscript{2}})$] through backward phase matching.
Since the length of the resonator is very long, the modes are spaced very closely, and hence, within a frequency range of interest, the acousto-optic coupling rate ($G_{ph}$) can be kept nearly constant.

\vspace{12pt}

Therefore, a longer SSH chain (Fig.~\ref{LongerChain}) can be realized in this new configuration, where the length of the chain is limited by the group index of the optical modes.
Furthermore, this system can also be extended for higher-order topological structures by incorporating more modes into the system.

\FloatBarrier

\section{Measurement of the Optical Reflection and Transmission}
\label{sec:MeasurementRefTrans}

\begin{figure}[th]
    \begin{adjustwidth}{-1in}{-1in}  
    \hsize=\linewidth
    \centering
    \includegraphics[width=0.8\textwidth]{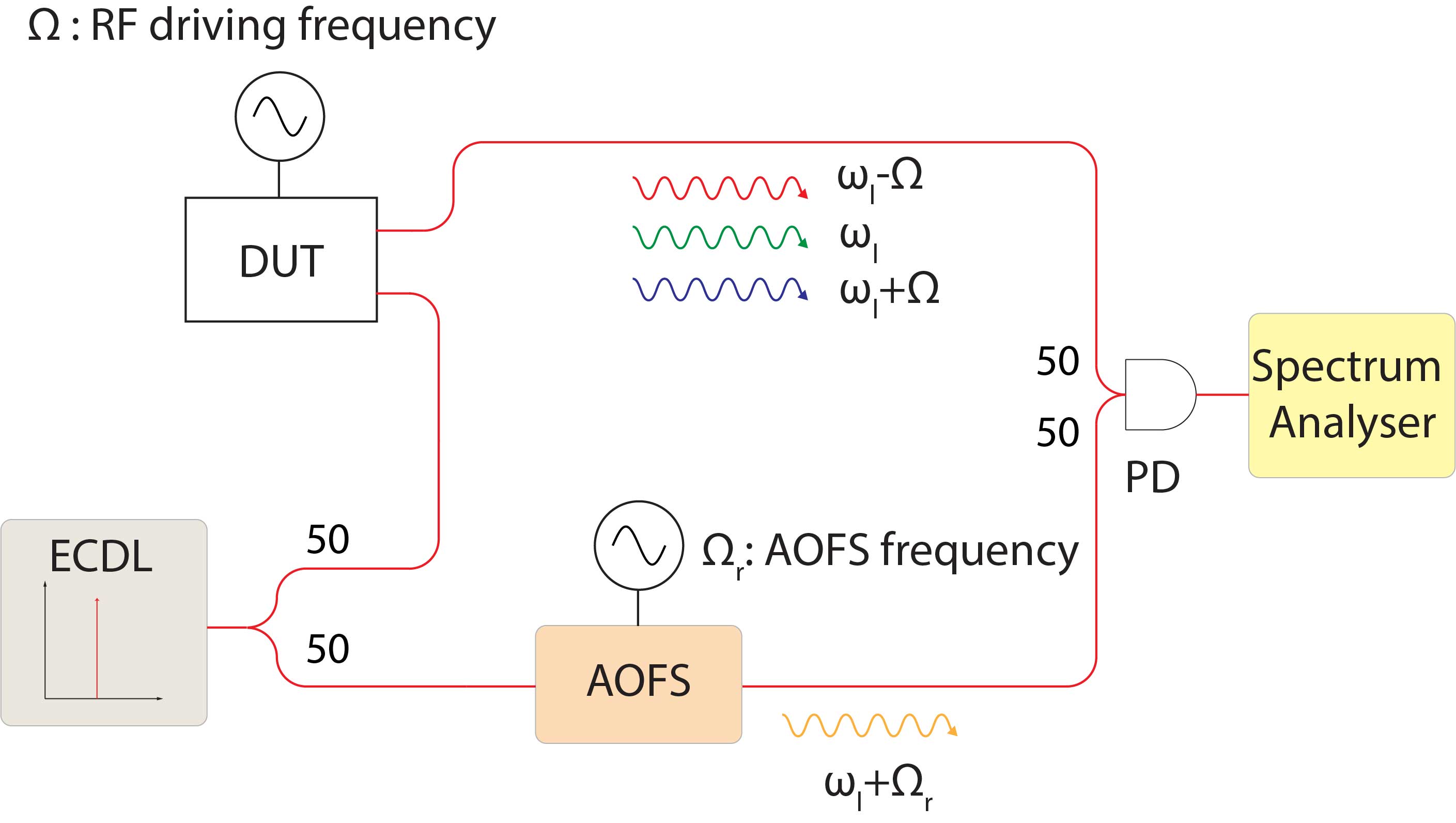}
    \caption{
    \textbf{Heterodyne detection system} ECDL: External cavity diode laser, AOFS: Acousto-optic frequency shifter, PD: Photodetector, DUT: Device under test
    }
    \label{supfig:2}
    \end{adjustwidth}
\end{figure}

In order to measure the Stokes, anti-Stokes, and carrier transmission, we utilize an optical heterodyne detection system by using an acousto-optic frequency shifter (Fig.~\ref{supfig:2})

Light is generated via an external cavity diode laser and split with a 50:50 coupler to realize two different optical paths. The source is a sub-50 kHz linewidth tunable external cavity diode laser (New Focus model TLB-6728-P). One of the arms probes the device under test while the other arm is used as a reference. An acousto-optic frequency shifter is used to offset the reference path by 100 MHz for heterodyne detection via a high-speed photodetector (PD). The directionality of the probing is controlled by an off-chip optical switch. Fiber polarization controllers (FPCs) are used to change the polarization of the light that is coupled to the chip.  

\vspace{12pt}

We can write the measured optical spectrum at the photo-detector as;

\begin{equation}
s_{out}=s_{out,0}e^{i\omega_lt}+s_{out,-1}e^{i(\omega_l-\Omega)t}+s_{out,+1}e^{i(\omega_l+\Omega)t}+s_{out,r}e^{i(\omega_l+\Omega_r)t}
\label{eq:optical_spectrum}
\end{equation}
The resulting electrical signals from the photodetector are the beat notes of the optical reference signal with the carrier, Stokes and anti-Stokes signals occur at $\Omega_r$, $\Omega+\Omega_r$ and $\Omega-\Omega_r$ respectively. Using equation \ref{eq:optical_spectrum}, we can find the RF outputs at the photodetector as,

\begin{equation}
P_{\Omega_r}=g_{pd}|s_{out,r}|^2|s_{out,0}|^2
\label{eq:AOFS_reference}
\end{equation}

\begin{equation}
P_{\Omega+\Omega_r}=g_{pd}|s_{out,r}|^2|s_{out,-1}|^2
\label{eq:Stokes_trans}
\end{equation}

\begin{equation}
P_{\Omega-\Omega_r}=g_{pd}|s_{out,r}|^2|s_{out,+1}|^2
\label{eq:antiStokes_trans}
\end{equation}
Where $P_{\Omega_r}$, $P_{\Omega+\Omega_r}$ and $P_{\Omega-\Omega_r}$  are the carriers, Stokes and anti-Stokes RF powers of the transmitted signals at the photodetector output. Here, we used $g_{pd}$ as a photodetector's optical to electrical conversion efficiency. Furthermore, the input carrier power to the system is also measured by the photodetector, and it is given by,

\begin{equation}
P_{in}=g_{pd}|s_{r}|^2|s_{in}|^2
\end{equation}
We can then normalize all the measured signals \ref{eq:AOFS_reference}-\ref{eq:antiStokes_trans} with respect to input power, and hence get;

\begin{equation}
\overline{P}_{\Omega_r}=|s_{out,0}/s_{in}|^2
\end{equation}

\begin{equation}
\overline{P}_{\Omega+\Omega_r}=|s_{out,-1}/s_{in}|^2
\end{equation}

\begin{equation}
\overline{P}_{\Omega-\Omega_r}=|s_{out,+1}/s_{in}|^2
\end{equation}

\begin{figure}[ht]
    \begin{adjustwidth}{-1in}{-1in}  
    \hsize=\linewidth
    \centering
    \includegraphics[width=0.8\textwidth]{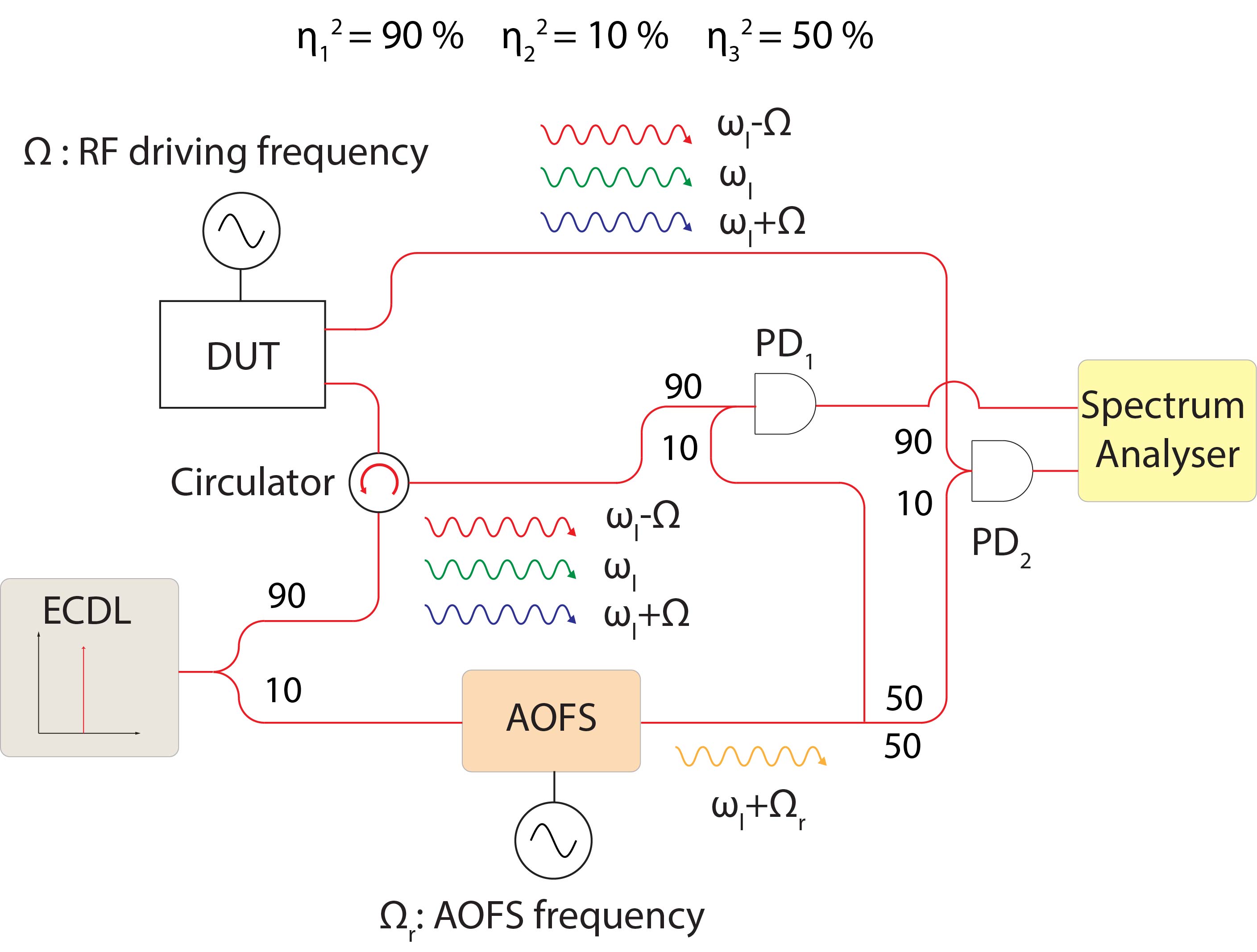}
    \caption{
    \textbf{Measurement setup that simultaneously measures the reflection and the transmission coefficients}
    The light from the ECDL is split into two reference arms with a single probe arm that has a circulator. Both reflected and transmitted signals are measured by the beat note signals that are obtained by the frequency-shifted AOFS signal.
    }
    \label{supfig:3}
    \end{adjustwidth}
\end{figure}

To measure the reflection signals alongside the transmission signals, we modify our experimental setup as in Fig.~\ref{supfig:3}. Here we assign the field transfer coefficients of the waveguide couplers as: $\eta_1$, $\eta_2$ and $\eta_3$ which is depicted in Fig.\ref{supfig:3}, and the laser output as $E$. At photodetector 1, we observe,

\begin{equation}
s_{out}^{1}=ER_{0}\eta_1^2e^{i\omega_lt}+ER_{-1}\eta_1^2e^{i(\omega_l-\Omega)t}+ER_{+1}\eta_1^2e^{i(\omega_l+\Omega)t}+E\eta_2^2\eta_3e^{i(\omega_l+\Omega_r)t}
\label{eq:detector_1}
\end{equation}
Here, $R\textsubscript{i}$ represents the reflection coefficient for the corresponding frequency component ($i$). Similarly, we can write the output spectrum at photodetector 2 as,

\begin{equation}
s_{out}^{2}=ET_0\eta_1^2e^{i\omega_lt}+ET_{-1}\eta_1^2e^{i(\omega_l-\Omega)t}+ET_{+1}\eta_1^2e^{i(\omega_l+\Omega)t}+E\eta_2^2\eta_3e^{i(\omega_l+\Omega_r)t}
\label{eq:detector_2}
\end{equation}
Here, $T\textsubscript{i}$ represents the transmission coefficient. The resulting electrical signals from the photodetectors are the beat notes of the optical reference signal with the carrier, Stokes, and anti-Stokes signals occurring at $\Omega_r$, $\Omega+\Omega_r$ and $\Omega-\Omega_r$, respectively. From equations \ref{eq:detector_1} and \ref{eq:detector_2}, we can find the RF outputs at both photodetectors as;

\begin{equation}
P_{1,\Omega_r}=g_{pd}|E^2\eta_1^2\eta_2^2\eta_3R_{0}^2|^2
\label{eq:AOFS_reference_detector_1}
\end{equation}

\begin{equation}
P_{1,\Omega+\Omega_r}=g_{pd}|E^2\eta_1^2\eta_2^2\eta_3R_{-1}^2|^2
\label{eq:Stokes_detector_1}
\end{equation}

\begin{equation}
P_{1,\Omega-\Omega_r}=g_{pd}|E^2\eta_1^2\eta_2^2\eta_3R_{+1}^2|^2
\label{eq:antiStokes_detector_1}
\end{equation}

\begin{equation}
P_{2,\Omega_r}=g_{pd}|E^2\eta_1^2\eta_2^2\eta_3T_{0}^2|^2
\label{eq:AOFS_reference_detector_2}
\end{equation}

\begin{equation}
P_{2,\Omega+\Omega_r}=g_{pd}|E^2\eta_1^2\eta_2^2\eta_3T_{-1}^2|^2
\label{eq:Stokes_detector_2}
\end{equation}

\begin{equation}
P_{2,\Omega-\Omega_r}=g_{pd}|E^2\eta_1^2\eta_2^2\eta_3T_{+1}^2|^2
\label{eq:antiStokes_detector_2}
\end{equation}
Where $P_{1,\Omega_r}$ ($P_{2,\Omega_r}$), $P_{1,\Omega+\Omega_r}$ ($P_{2,\Omega+\Omega_r}$), and $P_{1,\Omega-\Omega_r}$ ($P_{2,\Omega-\Omega_r}$) are the carrier, Stokes and anti-Stokes RF powers of the reflected (transmitted) signals at the photodetector output.

\vspace{12pt}

Furthermore, the input carrier power to the system is also measured by the photodetector, and it is given by,

\begin{equation}
P_{in}=g_{pd}|E^2\eta_1^2\eta_2^2\eta_3s_{in}^2|^2
\end{equation}

We can then normalize all the measured signals \ref{eq:AOFS_reference_detector_1}-\ref{eq:antiStokes_detector_2} with respect to input power, and get;

\begin{equation}
\overline{P}_{1,\Omega_r}=|R_{0}/s_{in}|^2
\label{eq:AOFS_normalized_detector_1}
\end{equation}

\begin{equation}
\overline{P}_{1,\Omega+\Omega_r}=|R_{-1}/s_{in}|^2
\label{eq:Stokes_normalized_detector_1}
\end{equation}

\begin{equation}
\overline{P}_{1,\Omega-\Omega_r}=|R_{+1}/s_{in}|^2
\label{eq:antiStokes_normalized_detector_1}
\end{equation}

\begin{equation}
\overline{P}_{2,\Omega_r}=|T_{0}/s_{in}|^2
\label{eq:AOFS_normalized_detector_2}
\end{equation}

\begin{equation}
\overline{P}_{2,\Omega+\Omega_r}=|T_{-1}/s_{in}|^2
\label{eq:Stokes_normalized_detector_2}
\end{equation}

\begin{equation}
\overline{P}_{2,\Omega-\Omega_r}=|T_{+1}/s_{in}|^2
\label{eq:antiStokes_normalized_detector_2}
\end{equation}
We then use these normalized signals \ref{eq:AOFS_normalized_detector_1}-\ref{eq:antiStokes_normalized_detector_2} to fit our data with our model.

\section{Calibration of the optical reflection}
\label{sec:Reflectioncalibration}

The measured reflected signal from the circulator involves the reflection due to optical components within the setup, such as fiber connectors, waveguide couplers, and v-groove, as well as the signal produced by the nanophotonic device. We need to separate these unwanted disturbances from our measured signal to fit our parameters successfully. Since these optical components (i.e., fiber connectors, waveguide couplers, and v-groove) do not show a dispersive behavior for narrowband measurements, they appear as a constant background that can be easily removed from the measured signal.

\begin{figure}[th]
    \begin{adjustwidth}{-1in}{-1in}   
    \hsize=\linewidth
    \includegraphics[width=1.2\textwidth]{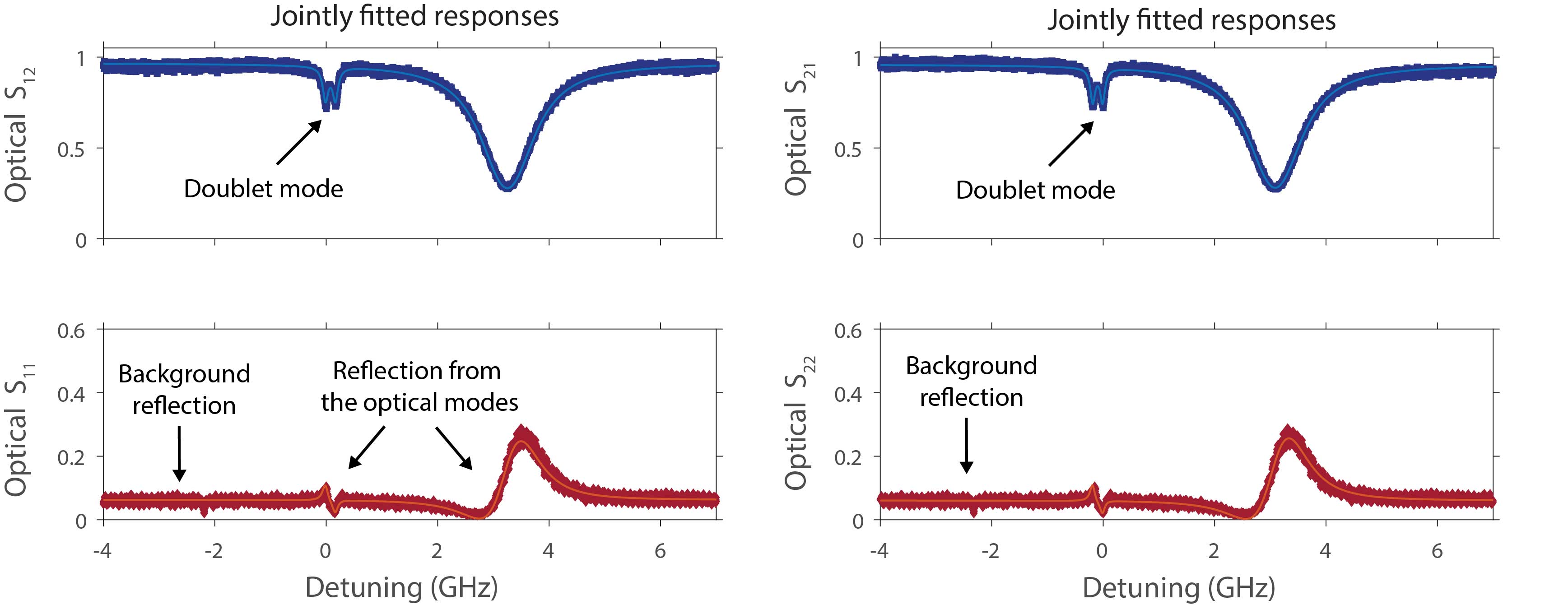}
    \centering
    \caption{
    \textbf{Simultaneously measured reflection and transmission data}
    The transmission data shows both optical modes (TE\textsubscript{00} and TE\textsubscript{10}). We also observe these modes as fano resonances in our reflection signal. This observation is due to the background reflection generated by other optical components in our experimental setup.
    }
    \label{supfig:5}
    \end{adjustwidth}
\end{figure}

\vspace{12pt}

To accomplish this task, we measure the transmitted and reflected signals from the device where the amplitude response is presented in Fig.~\ref{supfig:5}. Here, we center the optical detuning around TE\textsubscript{00} mode and plot the remaining experimental data for each figure. We observe two significant dips corresponding to the TE\textsubscript{00} and TE\textsubscript{10} modes at the transmitted signal. As expected, the TE\textsubscript{00} mode shows a doublet response since the Rayleigh scattering rate is comparable with its loss rate. However, the same observation is not realized for the TE\textsubscript{10} mode as this mode shows a larger loss rate. 
For the reflection measurement, we observe two fano-like resonances corresponding to the Rayleigh scattering from the optical resonators and a constant background reflection. We remove this background signal from our reflection measurement to extract the truly reflected signal due to the scatterers within these resonators.

\vspace{12pt}


\section{RF Electrode Characterization}
\label{sec:RF-S-Parameters}

\begin{figure}[th]
    \begin{adjustwidth}{-1in}{-1in}  
    \hsize=\linewidth
    \includegraphics[width=\textwidth]{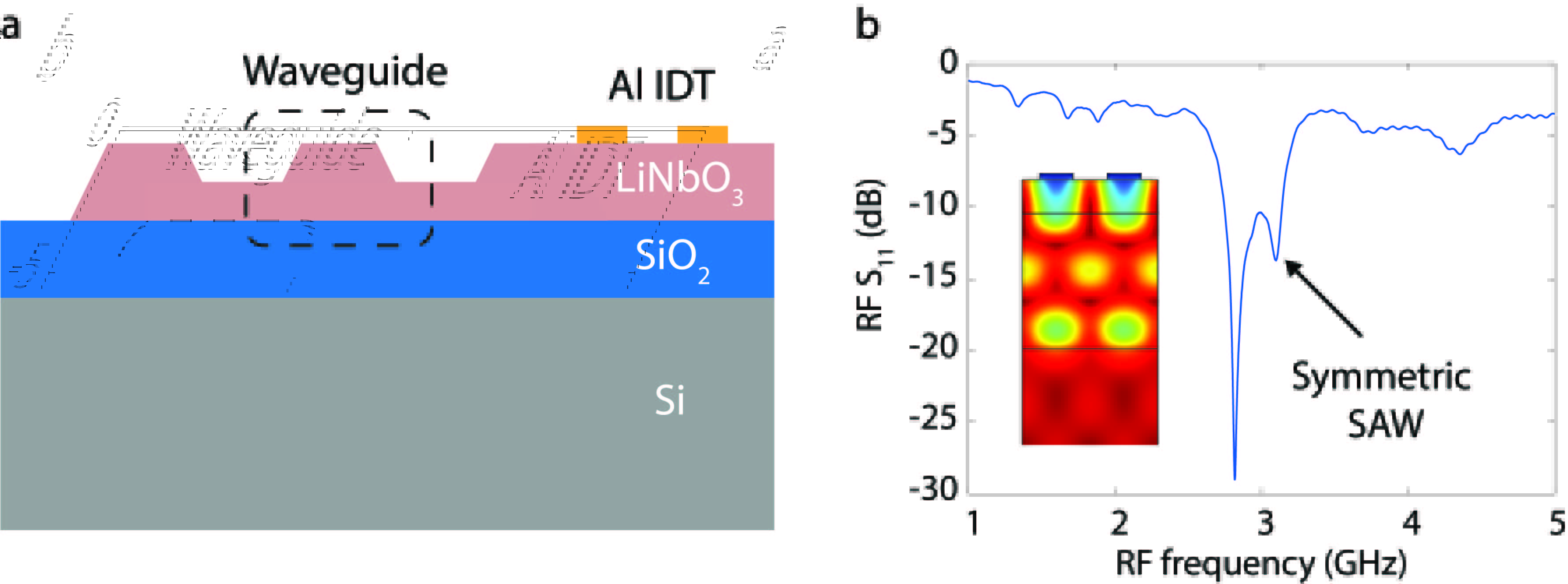}
    \centering
    \caption{
       \textbf{Characterization of the acoustic components in thin film LiNbO\textsubscript{3}} 
        \textbf{(a)} Schematic representation of the device cross-section shows the optical and acoustic components. The device is unreleased, and it works with the shown layers (LiNbO\textsubscript{3}-SiO\textsubscript{2}-Si)
        \textbf{(b)} IDT characterization using RF reflection measurement (RF $S_{11}$ parameter) shows that the surface acoustic wave in this device is efficiently generated around 3 GHz as determined by the designed IDT pitch and the surface acoustic wave speed. 
        }
    \label{supfig:6}
    \end{adjustwidth}
\end{figure}

We fabricated aluminum interdigital transducers (IDTs) to excite the required surface acoustic waves (SAW) in X-cut thin film LiNbO\textsubscript{3}. The cross-section of our material stack is given in Fig.~\ref{supfig:6}a, which is composed of 500 nm LiNbO\textsubscript{3}, 2 um SiO\textsubscript{3} and 500 um Si handle. Following our FEM simulations, we design and align our actuators along Y-30$^o$ to efficiently excite symmetric SAW modes in this stack. For experimental characterization, we use a calibrated VNA to measure the S\textsubscript{11} parameters of our IDT. The experimental results are given in Fig.~\ref{supfig:6}b. The reflection dip around 3 GHz corresponds to our designed symmetric SAW mode and experimentally shows a very large electromechanical coupling rate. We use this acoustic mode to induce strong chiral dispersion in our micro-resonators.

\begin{landscape}

\section{Summary of experimental parameters}
\label{sec:Systemparameters}

     \begin{centering}
		\begin{table}[h!]
		\small
			\caption{Parameters of our experimental demonstration of backscattering suppression}

			\begin{tabular}{ | >{\raggedright\arraybackslash}m{3cm} |  >{\raggedright\arraybackslash}m{6.5cm}  >{\raggedright\arraybackslash}m{1cm}  >{\raggedright\arraybackslash}m{1.8cm}   >{\raggedright\arraybackslash}m{1.8cm}  >{\raggedright\arraybackslash}m{1.8cm} | }
				
				\hline
				& Parameters & Unit & Device in Fig.~\ref{fig:3}a & Device in Fig.~\ref{fig:3}b & Device in Fig.~\ref{fig:4} \\ \hline \hline

                & $\Delta\omega$ of TE\textsubscript{10} and TE\textsubscript{00} modes ($\omega$\textsubscript{1}  - $\omega$\textsubscript{2}) & GHz & 3.06 & 3.06 & 3.18 \\ \cline{2-6}
                
				& Total loss rate of TE\textsubscript{00} mode ($\kappa$\textsubscript{2}) & GHz & 0.09 & 0.1 & 0.12 \\\cline{2-6}
				
			    & Total loss rate of TE\textsubscript{10} mode ($\kappa$\textsubscript{1}) & GHz & 1.01 & 1.01 & 1.08 \\\cline{2-6}
                & Quality factor of TE\textsubscript{00} mode (Q\textsubscript{2}) & - & 2.15 $\times$ 10\textsuperscript{6} & 1.93 $\times$ 10\textsuperscript{6} & 1.76 $\times$ 10\textsuperscript{6} \\\cline{2-6}
                Two-mode optical WGR & Quality factor of TE\textsubscript{10} mode (Q\textsubscript{1}) & - & 1.92 $\times$ 10\textsuperscript{5} & 1.92 $\times$ 10\textsuperscript{5} & 1.79 $\times$ 10\textsuperscript{5} \\\cline{2-6}
                & External coupling rate ($\kappa$\textsubscript{ex2}) & GHz & 0.02 & 0.01 & 0.01 \\\cline{2-6}
                & External coupling rate ($\kappa$\textsubscript{ex1}) & GHz &0.59 & 0.68 & 0.88 \\\cline{2-6}
                & Rayleigh scattering within TE\textsubscript{00} mode (V\textsubscript{2}) & GHz & $\approx$ 0 & 0.16  & 0.17 \\\cline{2-6}
                & Rayleigh scattering within TE\textsubscript{10} mode (V\textsubscript{1}) & GHz & $\approx$ 0& 0.21 &0.27 \\\cline{2-6}
                & Wave number difference ( $\vert$ k\textsubscript{1} - k\textsubscript{2} $\vert$) & m\textsuperscript{-1} &2.47 $\times$ 10\textsuperscript{5}& 2.47 $\times$ 10\textsuperscript{5}& 2.47 $\times$ 10\textsuperscript{5} \\\cline{2-6}
                & Center Wavelength & nm & 1526 & 1524 & 1556 \\\cline{2-6}
                
                \hline\hline
                
                & Transverse wave number (q\textsubscript{transverse}) & m\textsuperscript{-1} &2.84 $\times$ 10\textsuperscript{6}&2.84 $\times$ 10\textsuperscript{6}& 2.84 $\times$ 10\textsuperscript{6} \\\cline{2-6}
                & Propagating wave number (q\textsubscript{propagating}) & m\textsuperscript{-1} & 2.47 $\times$ 10\textsuperscript{5}& 2.47 $\times$ 10\textsuperscript{5}& 2.47 $\times$ 10\textsuperscript{5} \\\cline{2-6}
                & Total wave number (q\textsubscript{total}) & m\textsuperscript{-1} &2.85 $\times$ 10\textsuperscript{6}&2.85 $\times$ 10\textsuperscript{6}& 2.85 $\times$ 10\textsuperscript{6} \\\cline{2-6}
                Surface Acoustic Wave & IDT Pitch ($\lambda$) & $\mu$m & 2.2& 2.2 & 2.2 \\\cline{2-6}
                & IDT Aperture (W) & $\mu$m & 400 & 400 & 400 \\\cline{2-6}
                & IDT Angle ($\theta$) & degree & 4.98 & 4.98 & 4.98 \\\cline{2-6}
                & Center frequency ($\Omega$) & GHz & 3.06 & 3.06 & 3.02 \\\cline{2-6}
                 \hline\hline
                Acousto-optic Interaction & Phonon-enhanced optomechanical coupling (at 27 dBm applied RF power) & GHz & 0.68 & 0.66 & 0.87 \\\cline{2-6}
                \hline\hline
			
    \end{tabular}%

    \label{tab:DeviceParameters}
		\end{table}
	\end{centering}

\end{landscape}

{\footnotesize \putbib}
\end{bibunit}

\end{document}